# Cyber Epidemic Models with Dependences

Maochao Xu[1], Gaofeng Da[2] and Shouhuai Xu[3]

[1] Department of Mathematics, Illinois State University
mxu2@ilstu.edu
[2] Institute for Cyber Security, University of Texas at San Antonio
dagfvc@gmail.com
[3] Department of Computer Science, University of Texas at San Antonio
shxu@cs.utsa.edu (corresponding author)

**Abstract**

Studying models of cyber epidemics over arbitrary complex networks can deepen our understanding of cyber security from a whole-system perspective. In this paper, we initiate the investigation of cyber epidemic models that accommodate the *dependences* between the cyber attack events. Due to the notorious difficulty in dealing with such dependences, essentially all existing cyber epidemic models have assumed them away. Specifically, we introduce the idea of Copulas into cyber epidemic models for accommodating the dependences between the cyber attack events. We investigate the epidemic equilibrium thresholds as well as the bounds for both equilibrium and non-equilibrium infection probabilities. We further characterize the side-effects of assuming away the due dependences between the cyber attack events, by showing that the results thereof are unnecessarily restrictive or even incorrect.

## 1 Introduction

Cyberspace (or Internet) is perhaps the most complex man-made system. While cyberspace has become an indispensable part of the society, economy and national security, cyber attacks also have become an increasingly devastating problem. Despite studies and progresses in the past decades, our understanding of cyber security from a whole-system perspective, rather than from a component or building-block perspective, is still at its infant stage. This is caused by many factors, including the dearth of powerful mathematical models that can capture and reason the interactions between the cyber attacks and the cyber defenses.

Recently, researchers have started pursuing the cyber-security value of "biological epidemics"-like mathematical models. While conceptually attractive, biological epidemic models cannot be directly used to describe cyber security because there are many cyber-specific issues. One particular issue, which we initiate its study in the present paper, is the *dependences* between the cyber attack events. To the best of our knowledge, these dependences have been *explicitly* assumed away in essentially all existing cyber epidemic models, perhaps because they are notoriously difficult to cope with. Indeed, accommodating the dependences introduces yet another dimension of difficulty to cyber epidemic models that incorporate arbitrary complex network structures. However, the dependences are inherent because, for example, the events that computers get infected are not independent of each other, and a malware may first infect some computers because the users visit some malicious websites and then spread over the network. Moreover, cyber attacks may be well coordinated by intelligent malwares, and the coordination causes positive dependences between the attack events.

### 1.1 Our contributions

In this paper, we initiate the systematic study of a new sub-field in cyber epidemic models, namely understanding and characterizing the importance of the dependences between the attack events in cyber epidemic models that accommodate arbitrary complex network structures. This is demonstrated through a non-trivial generalization of the powerful push- and pull-based cyber epidemic model that was recently investigated in [19]. Specifically, we capture the dependences between the cyber attack events by incorporating the idea of Copulas into cyber epidemic models. To the best of our knowledge, this is the first systematic study of cyber epidemic models that accommodate dependences, rather than assuming them away. Specifically, we make two contributions.

First, we derive epidemic equilibrium thresholds, namely sufficient conditions under which the epidemic spreading enters a non-negative equilibrium (the spreading never dies out when there are pull-based attacks, meaning that only positive equilibrium is relevant under this circumstance). Some of the sufficient conditions are less restrictive but require hard-to-obtain information (i.e., these conditions are theoretically more interesting), and the others are more restrictive but require easy-to-obtain information (i.e., these conditions are practically more useful). We also derive bounds for the equilibrium infection probabilities and discuss their tightness. The bounds are easy to obtain/compute, and are useful especially when it is infeasible to obtain the equilibrium infection probabilities numerically (let alone analytically). For example, the upper bounds can be treated as the worst-case scenarios when provisioning defense resources. For Erdős-Rényi (ER) and power-law networks, we further propose to approximate the equilibrium infection probabilities by taking advantage of the bounds. The approximation results are smaller than the upper bounds and would not underestimate the number of infected nodes, meaning that the approximation results can lead to more cost-effective defense. We further present bounds for *non-equilibrium* infection probabilities, no matter whether the spreading converges to equilibrium or not. All the results are obtained by explicitly accommodating the dependence structures between the cyber attack events.

Second, we characterize the side-effects of assuming away the due dependences on the bounds for equilibrium infection probabilities, on the epidemic equilibrium thresholds, and on the non-equilibrium infection probabilities. We show that assuming away the due dependences can make the results thereof unnecessarily restrictive or even incorrect. We further discuss the cyber security implications of the side-effects.

It is worth mentioning that as a first step towards ultimately tackling the dependence problem in cyber epidemic models, the Copulas technique, which we use in the present paper, is appealing because of the following. On one hand, it leads to tractable models, while capable of coping with high-dimensional dependence (i.e., dependence between a large vector of random variables). On the other hand, there are families of copula structures that have been extensively investigated in the literature of Applied Probability Theory and Risk Management, and various methods have been developed for estimating the types and parameters of copula structures in practice. Of course, much research remains to be done before we can answer questions such as: What approach is the most appropriate for accommodating dependence in cyber epidemic models, under what circumstances?

## 1.2 Related work

Biological epidemic models can be traced back to McKendrick and Kermack [13, 10]. Such homogeneous biological epidemic models were introduced to computer science for characterizing the spreading of computer viruses in [9]. Heterogeneous epidemic models, especially the ones that accommodate arbitrary network structures, were not studied until recently [17, 6, 1]. These studies led to the full-fledged push- and pull-based cyber epidemic model [19], which is the starting point of the present paper. To the best of our knowledge, all existing cyber epidemic models, which aim to accommodate arbitrary network structures (including other recent studies such as [16, 11, 20] and the references therein), assumed that the attacks are *independent of each other*. This is plausible because accommodating arbitrary network structures in cyber epidemic models already make the resulting models difficult to analyze, and accommodating dependences introduces, as we show in the present paper, another dimension of difficulty to the models.

The only exception is due to our recent study [18], which is based on a different approach to modeling cyber epidemics [11]. The main contribution of [18] is to get rid of the exponential distribution assumptions for certain random variables. Moreover, the model in [18] can only accommodate the *specific* Marshall-Olkin dependence structure between the attack events. In contrast, we here accommodate *arbitrary* dependence structures between the attack events, while investigating the epidemic equilibrium thresholds, the equilibrium and non-equilibrium infection probabilities, and the side-effects of assuming away the due dependences. Many of these issues are not studied in the context of [18] because its focus is different. This explains why the present paper is the first systematic treatment of dependences in cyber epidemic models.

The dependence modeled in the present paper is static (i.e., time-invariant). This study inspired [5], which makes a further step towards modeling dynamic dependence between cyber attacks, but using a different modeling approach.

The rest of the paper is organized as follows. In Section 2 we briefly review some facts about Copulas. In Section 3 we investigate the generalized cyber epidemic model that accommodates the dependences between the cyber attack events. In Section 4 we characterize the side-effects as caused by assuming away the due dependences. In Section 5, we conclude the paper with future research problems.

The following table summarizes the main notations used in the paper.

| | |
|---|---|
| $\mathsf{G} = (\mathsf{V}, \mathsf{E})$ | the graph/network in which cyber epidemics occurs, where $\mathsf{V}$ is the node set and $\mathsf{E}$ is the edge set |
| $\deg(v)$ | the degree of node $v$ in graph $\mathsf{G} = (\mathsf{V}, \mathsf{E})$, which can be represented by adjacency matrix A |
| $\mathrm{I}_v(t), \mathrm{I}_{v,j}(t)$ | the state of node $v$ at time $t$: $\mathrm{I}_v(t) = 1$ means *infected* and 0 means *secure*; $\mathrm{I}_{v,j}(t)$ is the state of the $j$th neighbor of node $v$ (1 means *infected* and 0 means *secure*), where $1 \leq j \leq \deg(v)$ |
| $i_v(t)$ | the probability that node $v \in \mathsf{V}$ is *infected* at time $t$ |
| $i_{v,j}(t)$ | the condition probability that at time $t$ node $v \in \mathsf{V}$ is *secure* but the $j$th neighbor of node $v$ is *infected* |
| $i_v^-, i_v^+$ | lower and upper bounds for the non-equilibrium infection probability $\lim_{t \to \infty} i_v(t)$, where the system does *not* converge to any equilibrium |
| $i_v^*, i_{v,j}^*$ | the equilibrium infection probabilities that node $v$ and its $j$th neighbor are *infected*, respectively |
| $i^{*-}, i_v^{*+}$ | $i^{*-}$ is the lower bound for the equilibrium infection probability $i_v^*$ for *every* $v$, $i_v^{*+}$ is the upper bound for the equilibrium infection probability $i_v^*$ of node $v$ |
| $\mathbf{i}^*, \mathbf{i}_v^*$ | $\mathbf{i}^* = (i_1^*, \ldots, i_N^*)$ where $N = |\mathsf{V}|$, $\mathbf{i}_v^* = (i_{v,1}^*, \ldots, i_{v,\deg(v)}^*)$ |
| $\alpha$ | the probability that a secure node $v \in \mathsf{V}$ is infected by pull-based cyber attacks at a time step |
| $\beta$ | the probability that an infected node $v \in \mathsf{V}$ becomes secure at a time step |
| $\gamma$ | the probability that an infected node $u$ successfully attacks node $v$ over $(u, v) \in \mathsf{E}$ at a time step |
| $\rho(M)$ | spectral radius of matrix $M$ |
| $C_v$ | $\deg(v)$-copula describing the dependence between the push-based cyber attacks against node $v$ |
| $C$ | 2-copula describing the dependence between the pull-based attacks and the push-based attacks against a node |
| $\delta_C$ | the diagonal section of copula C, i.e., $\delta_C(u) = C(u, \ldots, u)$ |

## 2 Preliminaries

Copulas can model dependences by relating the individual marginal distributions to their multivariate joint distribution. In this paper we will use the $n$-copulas [8, 15]. Specifically, a function $\mathsf{C} : [0, 1]^n \mapsto [0, 1]$ is called $n$-copula if:

- $\mathsf{C}(u_1, \ldots, u_n)$ is increasing in each component $u_z$, $z \in \{1, \ldots, n\}$.

- $\mathsf{C}(u_1, \ldots, u_{z-1}, 0, u_{z+1}, \ldots, u_n) = 0$ for all $u_j \in [0, 1]$, $j = 1, \ldots, n$, $j \neq z$.

- $\mathsf{C}(1, \ldots, 1, u_z, 1, \ldots, 1) = u_z$ for all $u_z \in [0, 1]$, $z = 1, \ldots, n$.

- C is $n$-increasing, i.e., for all $(u_{1,1}, \ldots, u_{1,n})$ and $(u_{2,1}, \ldots, u_{2,n})$ in $[0, 1]^n$ with $u_{1,j} \leq u_{2,j}$ and for all $j = 1, \ldots, n$, it holds that

$$\sum_{z_1=1}^{2} \cdots \sum_{z_n=1}^{2} (-1)^{\sum_{j=1}^{n} z_j} \mathsf{C}(u_{z_1,1}, \ldots, u_{z_n,n}) \geq 0.$$

Let $R_1, \ldots, R_n$ be random variables with distribution functions $F_1, \ldots, F_n$, respectively. The joint distribution function is $F(r_1, \ldots, r_n) = \mathbb{P}(R_1 \leq r_1, \ldots, R_n \leq r_n)$. The well-known Sklar's theorem states that there exists an $n$-copula C such that $F(r_1, \ldots, r_n) = \mathsf{C}(F_1(r_1), \ldots, F_n(r_n))$.

There are many families of copulas [8, 15]. One example is the Gaussian copula with

$$C(u_1, \ldots, u_n) = \Phi_\Sigma \left( \Phi^{-1}(u_1), \ldots, \Phi^{-1}(u_n) \right),$$

where $\Phi^{-1}$ is the inverse cumulative distribution function of the standard normal distribution and $\Phi_\Sigma$ is the joint cumulative distribution function of a multivariate normal distribution with mean vector zero and covariance matrix equal to the correlation matrix $\Sigma$. For simplicity, we assume that the correlation matrix has the form

$$\Sigma = \begin{pmatrix} 1 & \sigma & \ldots & \sigma \\ \sigma & 1 & \sigma & \sigma \\ & & \ldots & \\ \sigma & \sigma & \ldots & 1 \end{pmatrix},$$

where $\sigma$ measures the correlation between two random variables. Therefore, the Gaussian copula can be rewritten as

$$C(u_1, \ldots, u_n) = \Phi_\sigma \left( \Phi^{-1}(u_1), \ldots, \Phi^{-1}(u_n) \right).$$

Another example is the Archimedean family with

$$C(u_1, \ldots, u_n) = \phi^{-1} \left( \phi(u_1) + \ldots + \phi(u_n) \right),$$

where function $\phi$ is called a generator of C and satisfies certain properties (see [14] for details). The Archimedean family contains many well-known copula functions such as the Clayton and Frank copulas [15, 3]. The generator of the Clayton copula is $\phi_\theta(u) = u^{-\theta} - 1$, and we have

$$C(u_1, \ldots, u_n) = \left[ \sum_{j=1}^{n} u_j^{-\theta} - n + 1 \right]^{-1/\theta}, \quad \theta > 0.$$

The generator of the Frank copula is $\psi_\xi(u) = \log\left(\frac{e^{-\xi u} - 1}{e^{-\xi} - 1}\right)$, and we have

$$C(u_1, \ldots, u_n) = -\frac{1}{\xi} \log \left\{ 1 + \frac{\prod_{j=1}^{n}(e^{-\xi u_j} - 1)}{(e^{-\xi} - 1)^{n-1}} \right\}, \quad \xi > 0.$$

For illustration purpose, we will use the Gaussian, Clayton and Frank copulas as examples.

In order to compare the effects of dependences, we need to compare the degrees of dependences. For this purpose, we use the *concordance* order [15, 8]. Let $C_1$ and $C_2$ be two copulas, we say $C_1$ is less than $C_2$ in concordance order if $C_1(u_1, \ldots, u_n) \leq C_2(u_1, \ldots, u_n)$ for all $0 \leq u_i \leq 1$, $i = 1, \ldots, n$. In particular, Gaussian copulas and Clayton copulas are increasing in $\sigma$ and $\theta$ in concordance order, respectively.

The following lemmas will be used in the paper.

**Lemma 1** ([15]) *Let C be any n-copula, then*

$$\max \left\{ \sum_{j=1}^{n} u_j - n + 1, 0 \right\} \leq C(u_1, \ldots, u_n) \leq \min\{u_1, \ldots, u_n\}.$$

**Lemma 2** ([15]) *Let C be an n-copula, then*

$$|C(u_1, \ldots, u_n) - C(v_1, \ldots, v_n)| \leq \sum_{j=1}^{n} |u_j - v_j|.$$

# 3 Cyber Epidemic Model With Arbitrary Dependences

Now we present and investigate the cyber epidemic model that accommodates the dependences between the cyber attack events. This is the first systematic treatment of dependences in cyber epidemic models.

## 3.1 The Model

As in [19], we consider an undirected finite network graph $\mathsf{G} = (\mathsf{V}, \mathsf{E})$, where $\mathsf{V} = \{1, 2, \ldots, N\}$ is the set of $N = |\mathsf{V}|$ nodes (vertices) that can abstract computers (or software components at an appropriate resolution), and $\mathsf{E} = \{(u, v) : u, v \in \mathsf{V}\}$ is the set of edges. Note that $\mathsf{G}$ abstracts the network structure according to which the push-based cyber attacks take place (e.g., malware spreading), where $(u, v) \in \mathsf{E}$ abstracts that node $u$ can attack node $v$. In both principle and practice, $\mathsf{G}$ can range from a complete graph (i.e., any $u \in \mathsf{V}$ can directly attack any $v \in \mathsf{V}$) to any specific graph structure (i.e., node $u$ may not be able to attack node $v$ directly because, for example, the traffic from node $u$ is filtered or $u$ is blacklisted by $v$), which explains why we should pursue general results without restricting the network/graph structures. Denote by $\mathsf{A} = (a_{vu})$ the adjacency matrix of $\mathsf{G}$, where $a_{vu} = 1$ if and only if $(u, v) \in \mathsf{E}$, and $a_{vu} = 0$ otherwise. Note that the problem setting naturally implies $a_{vv} = 0$. Denote by $\deg(v)$ the degree of node $v$. In a discrete-time model, node $v \in \mathsf{V}$ is either *secure* (but vulnerable to attacks) or *infected* (and can attack other nodes) at any time $t = 0, 1, \ldots$. At each time step, an infected node $v$ becomes secure with probability $\beta$, which abstracts the defense power. The model accommodates two large classes of cyber attacks: a secure node $v$ can become infected because of (i) pull-based cyber attacks with probability $\alpha$, which include drive-by-download attacks (i.e., node $v$ getting infected because its user visits a malicious website) and insider attacks (i.e., the user intentionally runs a malware on node $v$), or (ii) push-based cyber attacks launched by $v$'s infected neighbor $u$ over edge $(u, v) \in \mathsf{E}$ with probability $\gamma$.

Our extension to the above model is to accommodate the dependences between the push-based attacks as well as the dependences between the push-based attacks and the pull-based attacks. These attacks are not independent because the events that the nodes get infected are not independent of each other, and because the push-based attacks are not independent of the pull-based attacks (e.g., a malware could first infect some nodes via the pull-based cyber attacks and then launch the push-based cyber attacks from the infected nodes). Moreover, the dependences between the push-based attacks can model that intelligent malwares launch coordinated attacks against the secure nodes.

Specifically, let $\mathrm{I}_v(t)$ denote the state of node $v$ at time $t$, where $\mathrm{I}_v(t) = 1$ means $v$ is infected and 0 means $v$ is secure. Let $(\mathrm{I}_{v,1}(t), \ldots, \mathrm{I}_{v,\deg(v)}(t))$ denote the state vector of node $v$'s neighbors at time $t$, where

$$\mathrm{I}_{v,j}(t) = \begin{cases} 1, & \text{the } j\text{th neighbor of node } v \text{ is infected at time } t, \\ 0, & \text{otherwise.} \end{cases}$$

Define $i_v(t) = \mathbb{P}(\mathrm{I}_v(t) = 1)$ and $i_{v,j}(t) = \mathbb{P}(\mathrm{I}_{v,j}(t) = 1 | \mathrm{I}_v(t) = 0)$ where $j = 1, \ldots, \deg(v)$.

Let $X_v(t) = 1$ denote the event that node $v$ is infected at time $t+1$ because of the push-based cyber attacks, and $X_v(t) = 0$ otherwise. Let $X_{v,j}(t+1) = 1$ denote the event that node $v$ is infected at time $t+1$ by its $j$th neighbor, and $X_{v,j}(t+1) = 0$ otherwise. Note that $\mathbb{P}(X_{v,j}(t+1) = 1 | \mathrm{I}_v(t) = 0) = \gamma \cdot i_{v,j}(t)$. Since any dependence structure between $X_{v,1}(t+1), \ldots, X_{v,\deg(v)}(t+1)$ always can be accommodated by some copula function $C_v$, we have

$$\begin{aligned} & \mathbb{P}(X_v(t+1) = 0 | \mathrm{I}_v(t) = 0) \\ = & C_v \left( 1 - \mathbb{P}(X_{v,1}(t+1) = 1 | \mathrm{I}_v(t) = 0), \ldots, 1 - \mathbb{P}(X_{v,\deg(v)}(t+1) = 1 | \mathrm{I}_v(t) = 0) \right) \\ = & C_v \left( 1 - \gamma i_{v,1}(t), \ldots, 1 - \gamma i_{v,\deg(v)}(t) \right). \end{aligned} \quad (1)$$

Similarly, let $Y_v(t+1) = 1$ denote the event that node $v$ is infected at time $t+1$ because of the pull-based cyber attacks. Then, we have $\mathbb{P}(Y_v(t+1) = 1 | \mathrm{I}_v(t) = 0) = \alpha$. By further accommodating the dependence structure between the push-based attacks and the pull-based attacks via some copula function $C$, we have

$$\begin{aligned} & \mathbb{P}(\mathrm{I}_v(t+1) = 1 | \mathrm{I}_v(t) = 0) \\ = & 1 - \mathbb{P}(X_v(t+1) = 0, Y_v(t+1) = 0 | \mathrm{I}_v(t) = 0) \\ = & 1 - C \left( C_v \left( 1 - \gamma i_{v,1}(t), \ldots, 1 - \gamma i_{v,\deg(v)}(t) \right), 1 - \alpha \right). \end{aligned} \quad (2)$$

Note that
$$\mathbb{P}(I_v(t+1)=1|I_v(t)=1) = (1-\beta)i_v(t). \tag{3}$$

From Eqs. (1), (2) and (3), we obtain the probability that node $v \in \mathsf{V}$ is infected at time $t+1$ as:

$$\begin{aligned}
i_v(t+1) &= \mathbb{P}(I_v(t+1)=1) \\
&= \mathbb{P}(I_v(t+1)=1|I_v(t)=1)\mathbb{P}(I_v(t)=1) + \mathbb{P}(I_v(t+1)=1|I_v(t)=0)\mathbb{P}(I_v(t)=0) \\
&= (1-\beta) \cdot i_v(t) + \mathbb{P}(I_v(t+1)=1|I_v(t)=0) \cdot (1 - i_v(t)) \\
&= (1-\beta)i_v(t) + \left[1 - C\left(C_v\left(1 - \gamma i_{v,1}(t), \ldots, 1 - \gamma i_{v,\deg(v)}(t)\right), 1-\alpha\right)\right](1-i_v(t)).
\end{aligned} \tag{4}$$

We will analyze Eq. (4) for $v \in \mathsf{V}$ to characterize the effects of the dependence structures $C$ and $C_v$ and the side-effects of assuming them away. Note that for the special case that the $X_{v,j}$'s are independent of each other and the push-based attacks and the pull-based attacks are also independent of each other, Eq. (4) degenerates to the model in [19]. Note also that in order to characterize the side-effects of assuming away the dependences, we need to accommodate the dependences at a higher-level of abstraction than the model parameters $\alpha$ and $\gamma$. This is because the parameters are indeed relatively easier to obtain in experiments/practice (e.g., considering a single compromised neighbor that is launching the push-based attacks, and considering the pull-based attacks in the absence of the push-based attacks).

### 3.2 Epidemic Equilibrium Threshold and Bounds for Equilibrium Infection Probabilities

The concept of *epidemic equilibrium threshold* [19] naturally extends the well-known concept of *epidemic threshold* in that the former describes the condition under which the epidemic spreading converges to a non-negative equilibrium, whereas the latter traditionally describes the condition under which the epidemic spreading converges to 0 (i.e., the spreading dies out). Note that $\alpha > 0$ implies that the spreading will never die out and that $\alpha = 0$ is necessary for the spreading to die out. Denote by $i_v^*$ the equilibrium infection probability for node $v \in \mathsf{V}$. In the equilibrium, Eq. (4) becomes:

$$i_v^* = (1-\beta)i_v^* + \left[1 - C\left(C_v\left(1-\gamma i_{v,1}^*, \ldots, 1-\gamma i_{v,\deg(v)}^*\right), 1-\alpha\right)\right](1-i_v^*), \quad v \in \mathsf{V}. \tag{5}$$

In what follows, Theorem 1 gives a general epidemic equilibrium threshold (i.e., sufficient condition under which the spreading enters the equilibrium), and Theorem 2 gives a more succinct but more restrictive sufficient condition.

**Lemma 3** *Let* $\mathsf{A}$ *be the adjacency matrix of* $\mathsf{G}$. *If*

$$\rho(A) < \frac{(\beta+\alpha)^2}{\gamma\beta}, \tag{6}$$

*then system* (4) *has a unique equilibrium* $(i_1^*, \ldots, i_N^*) \in [0,1]^N$.

**Proof** For any $v \in \mathsf{V}$, define $f_v(\mathbf{x}) : [0,1]^N \to [0,1]$ as

$$f_v(\mathbf{x}) = \frac{1 - C\left(C_v\left(1-\gamma x_{v,1}, \ldots, 1-\gamma x_{v,\deg(v)}\right), 1-\alpha\right)}{\beta + 1 - C\left(C_v\left(1-\gamma x_{v,1}, \ldots, 1-\gamma x_{v,\deg(v)}\right), 1-\alpha\right)}, \quad v = 1, \ldots, N,$$

where $\mathbf{x} = (x_1, \ldots, x_N) \in [0,1]^N$. Define $\mathbf{f}(\cdot) : [0,1]^N \to [0,1]^N$, where $\mathbf{f}(\mathbf{x}) = (f_1(\mathbf{x}), \ldots, f_N(\mathbf{x}))$. According to the Banach fixed-point theorem [7], it is sufficient to show that $\mathbf{f}(\mathbf{x}) = \mathbf{x}$ has a unique solution $\mathbf{i}^*$; that is, we need to prove that $\mathbf{f}(\cdot)$ is a contraction mapping.

Let $\mathbf{x}, \mathbf{y} \in [0,1]^N$. Consider the distance between them in the Euclidean norm,

$$||\mathbf{f}(\mathbf{x}) - \mathbf{f}(\mathbf{y})|| = \sqrt{\sum_{v=1}^N (f_v(\mathbf{x}) - f_v(\mathbf{y}))^2} = \sqrt{\sum_{v=1}^N \left(\frac{\beta\Gamma_v}{\Delta_v}\right)^2},$$

where

$$\Gamma_v = C\left(C_v\left(1-\gamma x_{v,1},\ldots,1-\gamma x_{v,\deg(v)}\right),1-\alpha\right) - C\left(C_v\left(1-\gamma y_{v,1},\ldots,1-\gamma y_{v,\deg(v)}\right),1-\alpha\right),$$

$$\begin{aligned}\Delta_v &= \left(\beta+1-C\left(C_v\left(1-\gamma x_{v,1},\ldots,1-\gamma x_{v,\deg(v)}\right),1-\alpha\right)\right) \\ &\quad \cdot \left(\beta+1-C\left(C_v\left(1-\gamma y_{v,1},\ldots,1-\gamma y_{v,\deg(v)}\right),1-\alpha\right)\right).\end{aligned}$$

By Lemmas 1 and 2, it follows that

$$|\Gamma_v| \leq \gamma \sum_{k=1}^{\deg(v)} |x_{v,k} - y_{v,k}| \quad \text{and} \quad \Delta_v \geq (\beta+\alpha)^2.$$

Therefore, we have

$$||\mathbf{f}(\mathbf{x}) - \mathbf{f}(\mathbf{y})|| \leq \frac{\beta\gamma}{(\beta+\alpha)^2} \sqrt{\sum_{v=1}^{N}\left(\sum_{k=1}^{\deg(v)} |x_{v,k}-y_{v,k}|\right)^2}.$$

Moreover,

$$\begin{aligned}\sum_{v=1}^{N}\left(\sum_{k=1}^{\deg(v)} |x_{v,k}-y_{v,k}|\right)^2 &= (|x_1-y_1|,\ldots,|x_N-y_N|)A^2(|x_1-y_1|,\ldots,|x_N-y_N|)^T \\ &\leq ||(|x_1-y_1|,\ldots,|x_N-y_N|)||^2 ||A||^2 \\ &= ||\mathbf{x}-\mathbf{y}||^2 ||A||^2,\end{aligned}$$

where $||A||$ denotes the operator norm of $A$. Since $A$ is symmetric matrix, we have

$$||A|| = \rho(A).$$

From condition (6), it follows that

$$||\mathbf{f}(\mathbf{x}) - \mathbf{f}(\mathbf{y})|| \leq \frac{\beta\gamma\rho(A)}{(\beta+\alpha)^2}||\mathbf{x}-\mathbf{y}|| < ||\mathbf{x}-\mathbf{y}||,$$

which means that $\mathbf{f}(\cdot)$ is a contraction mapping. ∎

**Theorem 1** (general epidemic equilibrium threshold) *Let* $\mathsf{A}$ *be the adjacency matrix of* $\mathsf{G}$ *and* $\mathsf{D}$ *be the diagonal matrix with the $v$th ($1 \leq v \leq N$) diagonal element equal to*

$$h(\alpha,\beta,\gamma,\mathbf{i}_v^*) = \left|C\left(C_v\left(1-\gamma i_{v,1}^*,\ldots,1-\gamma i_{v,\deg(v)}^*\right),1-\alpha\right) - \beta\right|,$$

*where $i_v^*$ is the equilibrium infection probability that satisfies Eq. (5). Let $W = \mathsf{D}+\gamma\mathsf{A}$. If condition (6) holds, namely that system (4) has a unique equilibrium, and the spectral radius $\rho(W) < 1$, then $\lim_{t\to\infty} i_v(t) = i_v^*$ exponentially for all $v \in \mathsf{V}$.*

**Proof** According to Lemma 3, there is a unique solution for $i_v^*$ under condition (6). Denote by $r_v(t) = i_v(t) - i_v^*$. We want to identify a sufficient condition under which $\lim_{t\to\infty} |r_v(t)| = 0$ for all $v \in V$. Note that

$$\begin{aligned}r_v(t+1) &= r_v(t)\left[C\left(C_v\left(1-\gamma i_{v,1}^*,\ldots,1-\gamma i_{v,\deg(v)}^*\right),1-\alpha\right) - \beta\right] + (1-i_v(t)) \\ &\quad \times \left[C\left(C_v\left(1-\gamma i_{v,1}^*,\ldots,1-\gamma i_{v,\deg(v)}^*\right),1-\alpha\right) - C\left(C_v\left(1-\gamma i_{v,1}(t),\ldots,1-\gamma i_{v,\deg(v)}(t)\right),1-\alpha\right)\right].\end{aligned}$$

By Lemma 2, we have

$$\begin{aligned}
|r_v(t+1)| &\leq |r_v(t)|h(\alpha,\beta,\gamma,\mathbf{i}_v^*) + (1-i_v(t)) \\
&\quad \times \left|C_v\left(1-\gamma i_{v,1}^*,\ldots,1-\gamma i_{v,\deg(v)}^*\right) - C_v\left(1-\gamma i_{v,1}(t),\ldots,1-\gamma i_{v,\deg(v)}(t)\right)\right| \\
&\leq |r_v(t)|h(\alpha,\beta,\gamma,\mathbf{i}_v^*) + \gamma(1-i_v(t))\sum_{j=1}^{\deg(v)}|i_{v,j}^* - i_{v,j}(t)| \\
&\leq |r_v(t)|h(\alpha,\beta,\gamma,\mathbf{i}_v^*) + \gamma\sum_{j=1}^{\deg(v)}|r_{v,j}(t)|,
\end{aligned}$$

where

$$h(\alpha,\beta,\gamma,\mathbf{i}_v^*) = \left|C\left(C_v\left(1-\gamma i_{v,1}^*,\ldots,1-\gamma i_{v,\deg(v)}^*\right),1-\alpha\right) - \beta\right|.$$

Define

$$z_v(t+1) = z_v(t)h(\alpha,\beta,\gamma,\mathbf{i}_v^*) + \gamma\sum_{j=1}^{\deg(v)} z_{v,j}(t),$$

with $z_v(0) \equiv |r_v(0)|$ and $z_{v,j}(0) \equiv |r_{v,j}(0)|$ for $j = 1,\ldots,\deg(v)$. We see $|r_v(t)| \leq z_v(t)$ for any $t$. Let $\mathbf{z}(t) = (z_1(t),\ldots,z_n(t))^T$. Then, we have the following matrix form

$$\mathbf{z}(t+1) = W\mathbf{z}(t) = W^{t+1}\mathbf{z}(0), \tag{7}$$

where $W = D + \gamma A$, D is the diagonal matrix with diagonal element $h(\alpha,\beta,\gamma,\mathbf{i}_v^*)$, and A is the adjacency matrix of G. Since matrix $W$ is nonnegative and symmetric, the Spectral Theorem [12] says that $\rho(W)$ is real. By using the well-known Gelfand formula, if $\rho(W) < 1$, then $\lim_{t\to\infty} W^t = \mathbf{0}$ and therefore $\lim_{t\to\infty} \mathbf{z}(t) = \mathbf{0}$. Since

$$\rho(W) = \lim_{t\to\infty} \|W^t\|^{1/t} \quad \text{and} \quad \|W^t\| \sim [\rho(W)]^t, \quad t \to \infty,$$

where $\|\cdot\|$ is the norm in real space $\mathbb{R}^n$, we conclude that $\|W^t\|$ converges to 0 exponentially when $\rho(W) < 1$. This means that the convergence rate of $\lim_{t\to\infty} \mathbf{i}(t) = \mathbf{i}^*$ is at least exponential. ∎

Use of the sufficient condition given by Theorem 1 requires to know $\mathbf{i}^*$ (i.e., $i_v^*$ for all $v$), which is difficult to obtain analytically. It is therefore important to weaken this requirement. Now we present a sufficient condition that only requires the equilibrium infection probability $i_v^*$ for *some* $v$ (rather than for *all* $v \in \mathsf{V}$). According to [4], we have

$$\rho(W) \leq \max_{v \in \mathsf{V}} h(\alpha,\beta,\gamma,\mathbf{i}_v^*) + \gamma\rho(A).$$

Therefore, a more restrictive (than the one given by Theorem 1) sufficient condition is to require

$$\max_{v \in \mathsf{V}} h(\alpha,\beta,\gamma,\mathbf{i}_v^*) + \gamma\rho(A) < 1, \quad \text{namely} \quad \rho(A) < \frac{1 - \max_{v \in \mathsf{V}} h(\alpha,\beta,\gamma,\mathbf{i}_v^*)}{\gamma}.$$

According to Eq. (5), we have

$$h(\alpha,\beta,\gamma,\mathbf{i}_v^*) = \left|1 - \frac{\beta}{1-i_v^*}\right|.$$

Therefore, we obtain the following more restrictive, but more succinct, sufficient condition:

**Corollary 1** $\lim_{t\to\infty} i_v(t) = i_v^*$ *exponentially for all* $v \in \mathsf{V}$*, if*

$$\rho(A) \leq \min\left\{\frac{1 - \max_{v \in \mathsf{V}}|1-\beta/(1-i_v^*)|}{\gamma}, \frac{(\beta+\alpha)^2}{\gamma\beta}\right\}. \tag{8}$$

Applying the above sufficient condition still requires to know the minimal and maximal $i_v^*$'s, which is hard to obtain analytically. Although it is always possible to obtain them numerically, we would want to have some more general results without relying on numerical solutions. In what follows we present such a sufficient condition (Theorem 2), which requires the following Proposition 1 that presents bounds for the equilibrium infection probability. The bounds are certainly of independent value.

**Proposition 1** (bounds for equilibrium infection probabilities) *For any dependence structures $C$ and $C_v$, which may be unknown, the equilibrium infection probability $i_v^*$ for $v \in V$ satisfies $i^{*-} \leq i_v^* \leq i_v^{*+}$, where*

$$i^{*-} = \frac{\gamma - \beta}{\gamma} \mathrm{I}\{\gamma > \alpha + \beta\} + \frac{\alpha}{\beta + \alpha} \mathrm{I}\{\gamma \leq \alpha + \beta\}, \quad i_v^{*+} = \frac{\min\left\{\alpha + \frac{\gamma \deg(v)}{\beta+1}, 1\right\}}{\beta + \min\left\{\alpha + \frac{\gamma \deg(v)}{\beta+1}, 1\right\}}.$$

**Proof** Rewrite Eq. (5) as

$$i_v^* = \frac{1 - C\left(C_v\left(1 - \gamma i_{v,1}^*, \ldots, 1 - \gamma i_{v,\deg(v)}^*\right), 1 - \alpha\right)}{\beta + 1 - C\left(C_v\left(1 - \gamma i_{v,1}^*, \ldots, 1 - \gamma i_{v,\deg(v)}^*\right), 1 - \alpha\right)}. \quad (9)$$

By noticing the monotonicity in (9) and applying Lemma 1, we obtain

$$\frac{\max\left\{\gamma i_{v,1}^*, \ldots, \gamma i_{v,\deg(v)}^*, \alpha\right\}}{\beta + \max\left\{\gamma i_{v,1}^*, \ldots, \gamma i_{v,\deg(v)}^*, \alpha\right\}} \leq i_v^* \leq \frac{\min\left\{\alpha + \gamma \sum_{j=1}^{\deg(v)} i_{v,j}^*, 1\right\}}{\beta + \min\left\{\alpha + \gamma \sum_{j=1}^{\deg(v)} i_{v,j}^*, 1\right\}}. \quad (10)$$

Let us first consider the lower bound. Note that for each $v \in V$,

$$i_v^* \geq x_1 \stackrel{def}{=} \frac{\alpha}{\beta + \alpha}.$$

By substituting $x_1$ for $i_{v,j}^*$ in Ineq. (10), we have

$$i_v^* \geq x_2 \stackrel{\triangle}{=} \frac{\max\{\gamma x_1, \alpha\}}{\beta + \max\{\gamma x_1, \alpha\}}.$$

By substituting $x_2$ for $i_{v,j}^*$ in Ineq. (10), we obtain $x_3$. By repeating the substitution, we obtain a sequence $\{x_n, n \geq 1\}$ with

$$x_n = \frac{\max\{\gamma x_{n-1}, \alpha\}}{\beta + \max\{\gamma x_{n-1}, \alpha\}}, \quad x_0 = 0.$$

Since $\{x_n, n \geq 1\}$ is increasing and bounded, we can get its limit, namely $i^{*-}$, by solving the following equation

$$\frac{\max\{\gamma x, \alpha\}}{\beta + \max\{\gamma x, \alpha\}} = x.$$

For the upper bound, note that $i_v^* \leq \frac{1}{\beta+1}$ for $v \in V$. By substituting $1/(\beta+1)$ for $i_{v,j}^*$ in Ineq. (10), we get $i_v^{*+}$. ∎

It is useful to know when the bounds in Proposition 1 are tight. For this purpose, we observe that if $\frac{\gamma}{\beta+1} \deg(v) \approx 0$, meaning that $\gamma \deg(v) << 1$ and that the attack-power is not strong, we have $i_v^{*+} \approx i^{*-} = \frac{\alpha}{\beta + \alpha}$. This means that the bounds are tight when the attack-power is not strong. On the other hand, Proposition 1 allows us to derive the following more succinct, but more restrictive (than Corollary 1 and therefore Theorem 1), sufficient condition for the epidemic spreading converges to the equilibrium (i.e., epidemic equilibrium threshold). The new sufficient condition involves the bounds $i^{*-}$ and $i_v^{*+}$ only (i.e., none of the equilibrium probabilities that are hard to obtain analytically).

**Theorem 2** (succinct epidemic equilibrium threshold) *The spreading enters the unique equilibrium if*

$$\rho(A) \leq \frac{1 - \max_{v \in \mathsf{V}} \left\{ \max \left\{ \left|1 - \frac{\beta}{1 - i^{*-}}\right|, \left|1 - \frac{\beta}{1 - i_v^{*+}}\right| \right\} \right\}}{\gamma},$$

*where $i^{*-}$ and $i_v^{*+}$ are defined in Proposition 1.*

**Proof** Note that for any $v \in \mathsf{V}$, we have

$$\max_{v \in \mathsf{V}} h(\alpha, \beta, \gamma, \mathbf{i}_v^*) = \max_{v \in \mathsf{V}} \left|1 - \frac{\beta}{1 - i_v^*}\right| \leq \max_{v \in \mathsf{V}} \left\{ \max \left\{ \left|1 - \frac{\beta}{1 - i^{*-}}\right|, \left|1 - \frac{\beta}{1 - i_v^{*+}}\right| \right\} \right\}.$$

Note that

$$\left|1 - \frac{\beta}{1 - i^{*-}}\right| \geq 1 - \frac{(\beta + \alpha)^2}{\beta},$$

which implies

$$\max_{v \in \mathsf{V}} \left\{ \max \left\{ \left|1 - \frac{\beta}{1 - i^{*-}}\right|, \left|1 - \frac{\beta}{1 - i_v^{*+}}\right| \right\} \right\} \geq 1 - \frac{(\beta + \alpha)^2}{\beta}.$$

Therefore,

$$1 - \max_{v \in \mathsf{V}} \left\{ \max \left\{ \left|1 - \frac{\beta}{1 - i^{*-}}\right|, \left|1 - \frac{\beta}{1 - i_v^{*+}}\right| \right\} \right\} \leq \min \left\{ 1 - \max_{v \in V} |1 - \beta/(1 - i_v^*)|, \frac{(\beta + \alpha)^2}{\beta} \right\}.$$

According to Corollary 1, we obtain the desired result. ∎

### 3.3 Tighter Bounds for Equilibrium Infection Probabilities in Star and Regular Networks

**Star networks.** A star-shaped network consists of a hub and $(N - 1)$ leaves that are connected only to the hub. Hence, the adjacency matrix A can be represented as

$$A = \begin{pmatrix} 0 & 1 & \ldots & 1 \\ 1 & 0 & \ldots & 0 \\ \vdots & \vdots & \ldots & 0 \\ 1 & 0 & \ldots & 0 \end{pmatrix}_{N \times N}$$

The spectral radius is $\rho(A) = \sqrt{N - 1}$. In this case, Eq. (5) becomes:

$$i_h^* = \frac{1 - C\left(\delta_{C_h}\left(1 - \gamma i_l^*\right), 1 - \alpha\right)}{\beta + 1 - C\left(\delta_{C_h}\left(1 - \gamma i_l^*\right), 1 - \alpha\right)}, \tag{11}$$

$$i_l^* = \frac{1 - C\left(1 - \gamma i_h^*, 1 - \alpha\right)}{1 + \beta - C\left(1 - \gamma i_h^*, 1 - \alpha\right)}, \tag{12}$$

where $i_h^*$ and $i_l^*$ are the equilibrium probabilities that the hub and the leaves are infected, respectively. Note that the effect of the copula $C_h$ on the equilibrium probabilities only depends on its diagonal section $\delta_{C_h}$.

In what follows we present two results about the equilibrium infection probabilities, which are not implied by the above general results that apply to arbitrary network structures. First, we can prove $i_h^* \geq i_l^*$.

**Proposition 2** *For the star networks, it holds that $i_h^* \geq i_l^*$.*

**Proof** Denote by

$$f(x) = \frac{1 - C\left(\delta_{C_h}\left(1 - \gamma x\right), 1 - \alpha\right)}{\beta + 1 - C\left(\delta_{C_h}\left(1 - \gamma x\right), 1 - \alpha\right)} \quad \text{and} \quad g(x) = \frac{1 - C\left(1 - \gamma x, 1 - \alpha\right)}{1 + \beta - C\left(1 - \gamma x, 1 - \alpha\right)}$$

where $x \in [0,1]$. Since $\delta_{C_h}(x) \leq x$, we have $f(x) \geq g(x)$. Suppose $i_h^* < i_l^*$ and $(i_h^*, i_l^*)$ is a solution to Eqs. (11) and (12). Then, $i_h^* = f(i_l^*) = g^{-1}(i_l^*) < i_l^*$. Since $g(x)$ is increasing in $x$ and so is $g^{-1}$, we have $i_l^* \leq g(i_l^*)$ and $f(i_l^*) < i_l^* \leq g(i_l^*)$, which contradicts with $f(x) \geq g(x)$ for $x \in [0,1]$. ∎

Second, we present refined bounds for equilibrium infection probabilities $i_h^*$ and $i_l^*$. The bounds are useful because even in the case of star networks, it is hard to derive analytic expressions and infeasible to numerically compute (especially for complex dependence structures) $i_h^*$ and $i_l^*$.

**Proposition 3** (tighter upper bounds for the equilibrium infection probabilities in star networks) *For star networks and regardless of the dependence structures (which can be unknown), we have $i^{*-} \leq i_h^* \leq i_h^{*+}$ and $i^{*-} \leq i_l^* \leq i_l^{*+}$, where $i^{*-}$ is defined in Proposition 1 and*

$$i_h^{*+} = \frac{1}{\beta + 1} \mathrm{I}\left\{\frac{1}{\beta + 1} \geq \frac{1 - \alpha}{(N-1)\gamma}\right\}$$
$$+ \frac{(N-1)\gamma - \alpha - \beta + \sqrt{((N-1)\gamma - \alpha - \beta)^2 + 4(N-1)\gamma\alpha}}{2(N-1)\gamma} \mathrm{I}\left\{\frac{1}{\beta + 1} < \frac{1 - \alpha}{(N-1)\gamma}\right\}.$$

*and*

$$i_l^{*+} = \frac{1}{\beta + 1} \mathrm{I}\left\{\frac{1}{\beta + 1} \geq \frac{1 - \alpha}{\gamma}\right\} + \frac{\gamma - \alpha - \beta + \sqrt{(\gamma - \alpha - \beta)^2 + 4\gamma\alpha}}{2\gamma} \mathrm{I}\left\{\frac{1}{\beta + 1} < \frac{1 - \alpha}{\gamma}\right\}.$$

**Proof** The lower bound $i^{*-}$ is the same as in Proposition 1. Let's focus on $i_h^{*+}$. From Ineq. (10), we have

$$i_h^* \leq \frac{\min\{\alpha + (N-1)\gamma i_l^*, 1\}}{\beta + \min\{\alpha + (N-1)\gamma i_l^*, 1\}} \triangleq f(i_l^*).$$

Since the right-hand side of the above inequality increases in $i_l^*$, by Proposition 2 we have

$$i_h^* \leq f(i_h^*),$$

and therefore $i_h^* \leq i_h^{*+}$, where $i_h^{*+}$ is the solution to equation

$$x = f(x). \tag{13}$$

For the upper bound $i_l^{*+}$, we can similarly obtain the desired result by solving equation

$$x = f\left(\frac{x}{N-1}\right). \tag{14}$$

∎

Now we explain why the upper bounds $i_h^{*+}$ and $i_l^{*+}$ given by Proposition 3 are smaller (i.e., tighter) than the general upper bounds that can be derived from Proposition 1 by instantiating $\mathsf{G} = (\mathsf{V}, \mathsf{E})$ as star networks. To see this, we note that $i_h^{*+}$ is the solution to Eq. (13) and $i_h^{*+} \leq \frac{1}{\beta+1}$, meaning that

$$i_h^{*+} \leq \frac{\min\left\{\alpha + (N-1)\frac{\gamma}{\beta+1}, 1\right\}}{\beta + \min\left\{\alpha + (N-1)\frac{\gamma}{\beta+1}, 1\right\}},$$

where the right-hand side of the inequality is exactly the upper bound that can be derived from Proposition 1 by substituting $\deg(v)$ with the degree of the hub. This means that $i_h^{*+}$ is smaller than the upper bound given by Proposition

1. Similarly, we can show that $i_l^{*+}$ is smaller than the upper bound given by Proposition 1. Moreover, by comparing (13) and (14), we see that $i_h^{*+} \geq i_l^{*+}$. Since the lower bound $i^{*-}$ is the same as the lower bound given by Proposition 1, we conclude that the bounds given by Proposition 3 are tighter than the bounds given by Proposition 1.

To see the tightness of the bounds given by Proposition 3, we consider two combinations of dependence structures: $(C, C_v)$=(Gaussian,Frank) and $(C, C_v)$=(Gaussian, Clayton) with parameters $\sigma = \theta = \xi = 0.1$ as reviewed in Section 2. Figure 1 plots $i_h^*$, $i_l^*$, $i^{*-}$, $i_h^{*+}$, and $i_l^{*+}$ for $N = 3, \ldots, 81$ with $(\alpha, \beta, \gamma) = (0.5, 0.1, 0.1)$; all these parameter settings satisfy condition (8). We observe that the upper bound $i_h^{*+}$ becomes flat for $N \geq 5$, because it causes $i_h^{*+} = \frac{1}{\beta+1}$ (i.e., independent of $N$); whereas, the upper bound $i_l^{*+}$ is flat because it is always independent of $N$. We observe that the upper bound for hub node, $i_h^{*+}$, becomes extremely tight for dense star networks with $N > 40$. However, the upper bound for leave nodes almost always exhibits that $i_l^{*+} - i_l^* \approx 0.011$ (i.e., the upper bound overestimates about 0.88 infected nodes for a star network of $N = 80$ nodes). In any case, the upper bounds only somewhat overestimate the numerical solutions $i_v^*$'s and thus can be used for decision-making purpose when $i_v^*$'s are infeasible to compute.

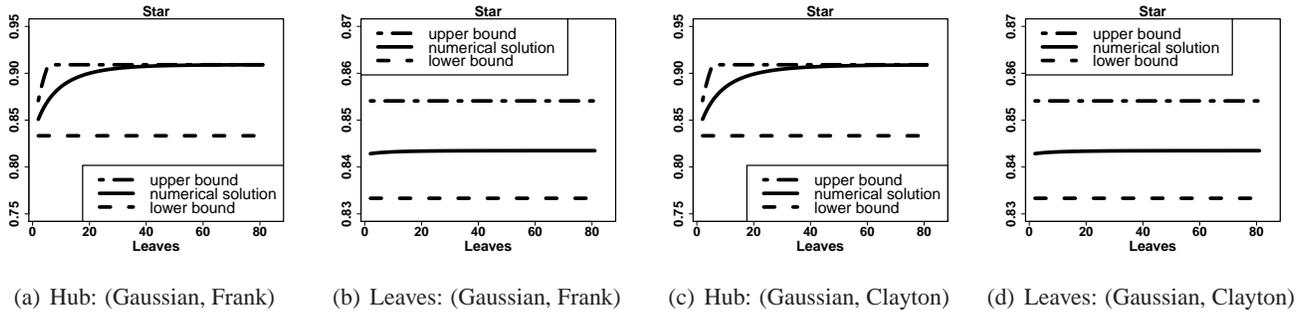

(a) Hub: (Gaussian, Frank)    (b) Leaves: (Gaussian, Frank)    (c) Hub: (Gaussian, Clayton)    (d) Leaves: (Gaussian, Clayton)

Figure 1: Star networks: upper bound $i_h^{*+}$ for hub ($i_l^{*+}$ for leaves) vs. numerical solution $i_h^*$ for hub ($i_l^*$ for leaves) vs. lower bound $i^{*-}$ (for both hub and leaves) with respect to $(\alpha, \beta, \gamma) = (0.5, 0.1, 0.1)$ and $(C, C_v)$.

**Regular networks.** For regular networks, each node $v \in \mathsf{V}$ has degree $d$ for some $d \in [1, N-1]$ and $\rho(\mathsf{A}) = d$. According to Proposition 1, we have

$$i_v^* = \frac{1 - C\left(C_v\left(1 - \gamma i_{v,1}^*, \ldots, 1 - \gamma i_{v,d}^*\right), 1 - \alpha\right)}{\beta + 1 - C\left(C_v\left(1 - \gamma i_{v,1}^*, \ldots, 1 - \gamma i_{v,d}^*\right), 1 - \alpha\right)}, \quad v \in \mathsf{V}.$$

Now we want to present refined bounds for equilibrium infection probability $i_v^*$.

**Proposition 4** *(tighter upper bound for the equilibrium infection probability in regular networks) For regular network $G = (\mathsf{V}, \mathsf{E})$ and regardless of the dependence structures (which can be unknown), we have $i^{*-} \leq i_v^* \leq i^{*+}$ for any $v \in \mathsf{V}$, where $i^{*-}$ is defined in Proposition 1 and*

$$i^{*+} = \frac{1}{\beta+1}\mathrm{I}\left\{\frac{1}{\beta+1} \geq \frac{1-\alpha}{\gamma d}\right\} + \frac{\gamma d - \alpha - \beta + \sqrt{(\gamma d - \alpha - \beta)^2 + 4\gamma \alpha d}}{2\gamma d}\mathrm{I}\left\{\frac{1}{\beta+1} < \frac{1-\alpha}{\gamma d}\right\}.$$

**Proof** Define function
$$f(x) = \frac{\min\{\alpha + \gamma dx, 1\}}{\beta + \min\{\alpha + \gamma dx, 1\}}$$

and a sequence $\{x_n, n \geq 0\}$ with $x_n = f(x_{n-1})$, $x_0 = 1/(\beta+1)$. Observe that for all $v \in \mathsf{V}$, we have $i_v^* \leq x_0$ and hence from Ineq. (10), it follows that $i_v^* \leq x_1$ for all $v \in \mathsf{V}$. By repeating this process, we have $i_v^* \leq x_n$ for all $n$. Since $f(x)$ is increasing and $x_1 \leq x_0$, $x_n$ is decreasing in $n$. Thus, we have $i_v^* \leq i^{*+}$, which is the solution of the equation $x = f(x)$. If $\frac{1}{\beta+1} \geq \frac{1-\alpha}{\gamma d}$, then $i^{*+} = \frac{1}{\beta+1}$; otherwise, $i^{*+}$ is the positive solution to equation $\gamma d x^2 + (\alpha + \beta - \gamma d)x - \alpha = 0$. Thus, we obtain the desired result. ∎

Note that the upper bound $i^{*+}$ given by Proposition 4 is smaller than the upper bound $i_v^{*+}$ obtained by instantiating $\deg(v) = d$ in Proposition 1, because $i_v^{*+}$ is exactly the $x_1$ defined in the proof of Proposition 4. To see the tightness of bounds $i^{*-}$ and $i^{*+}$ given by Proposition 4, we consider $(C, C_v)$=(Gaussian,Frank) and $(C, C_v)$=(Gaussian, Clayton) with parameters $\sigma = \theta = \xi = 0.1$ as reviewed in Section 2. Figure 2 plots numerical $i_v^*$, $i^{*-}$ and $i^{*+}$ with respect to node degree $d = 2, \ldots, 80$ with $(\alpha, \beta, \gamma) = (0.5, 0.1, 0.01)$; all these parameter settings satisfy condition (8). We observe that $i_v^{*+}$ becomes flat for sufficiently dense regular networks. This is because $i_v^{*+} = \frac{1}{\beta+1}$ when $d \geq \frac{(1-\alpha)(\beta+1)}{\gamma}$. For $(C, C_v)$=(Gaussian,Frank), we further observe that the upper bound $i_v^{*+}$ is reasonably tight especially for relatively sparse regular networks, with $i_v^{*+} - i_v^* < 0.021$ for $d < 20$ (i.e., for a sparse regular network of $N = 1000$ nodes, the upper bound only overestimates at most 21 infected nodes). Even for dense regular network with $d > 20$, we have $i_v^{*+} - i_v^* \leq 0.038$ (i.e., for a dense regular network of $N = 1000$ nodes, the upper bound only overestimates at most 38 infected nodes), where equality holds for $d = 54$. For $(C, C_v)$=(Gaussian, Clayton), we also observe that the upper bound $i_v^{*+}$ is tight especially for relatively sparse regular networks with $d < 20$ and $i_v^{*+} - i_v^* < 0.021$ (i.e., for a sparse regular network of $N = 1000$ nodes, the upper bound only overestimates at most 21 infected nodes). Even for dense regular network with $d > 20$, we have $i_v^{*+} - i_v^* \leq 0.039$, where equality holds for $d = 54$. This means that for decision-making purpose, the defender can use the upper bound $i_v^{*+}$ instead of the numerical solution $i_v^*$, especially when $i_v^*$ is infeasible to compute.

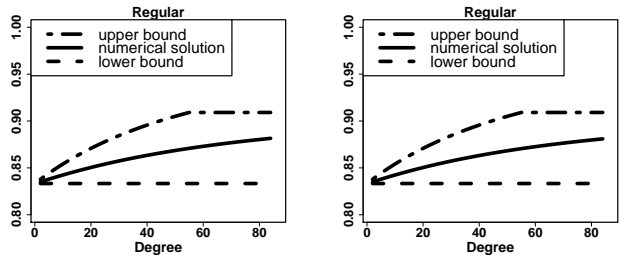

(a) (Gaussian,Frank,0.5,0.1,0.01)  (b) (Gaussian,Clayton,0.5,0.1,0.01)

Figure 2: Regular networks: upper bound $i_v^{*+}$ vs. numerical solution $i_v^*$ vs. lower bound $i_v^{*-}$ with respect to $(C, C_v, \alpha, \beta, \gamma)$

### 3.4 Approximating Equilibrium Infection Probabilities in ER and Power-law Networks

For star and regular networks, we have derived tighter bounds for equilibrium infection probabilities (than the general bounds given by Proposition 1). Unfortunately, we do not know how to derive tighter bounds for ER and power-law networks. As an alternative, we propose to approximate equilibrium infection probabilities by taking advantage of the upper and lower bounds. The approximation is useful because it is often smaller than the upper bound, which never underestimates, but may substantially overestimate, the threats in terms of equilibrium infection probabilities. That is, the approximation method can lead to more cost-effective defense than the upper bound.

The approximation method is the following: We first compute lower bounds, upper bounds, and numerical solutions for a feasible number of instances of $(\mathsf{G}, C, C_v, \alpha, \beta, \gamma)$, based on given computer resources. We then use the resulting data to derive (via statistical methods) some function of the lower and upper bounds. For even larger $\mathsf{G}$ of the same type as well as $(C, C_v)$ of the same kind, the resulting function would be smaller than the upper bound and would not underestimate the equilibrium infection probabilities. The key insight is that we can compute, for networks of *any* size, the upper and lower bounds according to Proposition 1. This means that we can approximate the equilibrium infection probabilities for arbitrarily large networks, for which it is often infeasible to numerically (let alone analytically) compute the equilibrium infection probabilities.

To illustrate the approximation method, we also consider $(C, C_v)$=(Gaussian,Frank) and $(C, C_v)$=(Gaussian, Clayton) with parameters $\sigma = \theta = \xi = 0.1$ as reviewed in Section 2. We use the `erdos.renyi.game` generator of the `igraph` package in the R system to generate a random ER network of $N = 1000$ nodes and edge probability 0.01; the resulting network instance has spectral radius 11.38045. We use the `static.power.law.game` generator of the `igraph` package in the R system to generate a random power-law network of $N = 1000$ nodes, 5000 edges, and power-law exponent 2.1 (note that 2.1 is the power-law exponent of the Internet AS-level network [6]); the resulting network instance has spectral radius 22.97582. We consider combinations of $(\alpha, \beta, \gamma)$ that satisfy condition (8), where $\alpha \in \{0.01, 0.05, 0.1, 0.2, 0.3, 0.4, 0.5\}$, $\beta \in \{0.1, 0.2, 0.3, 0.4, 0.5, 0.6, 0.7, 0.8, 0.9\}$, $\gamma \in \{0.01, 0.02, 0.03, 0.04, 0.05, 0.06, 0.07, 0.08, 0.09, 0.1\}$. It turns out that for $(C, C_v)$=(Gaussian, Frank), the ER network has 307 combinations of $(\alpha, \beta, \gamma)$ that satisfy condition (8); the power-law network has 125 combinations of $(\alpha, \beta, \gamma)$ that satisfy condition (8), because the spectral radius is larger. For $(C, C_v)$=(Gaussian, Clayton), the ER network has 307 combinations of $(\alpha, \beta, \gamma)$ that satisfy condition (8); the power-law network has 126 combinations of $(\alpha, \beta, \gamma)$ that satisfy condition (8).

We compute equilibrium infection probability $i_v^*$ numerically by solving Eqs. (5) for $v \in \mathsf{V}$ via the `BB` package in the R system. We compute the upper and lower bounds, namely $i^{*-}$ and $i_v^{*+}$, according to Proposition 1. Since it is infeasible to numerically compute $i_v^*$ for large networks, we propose to approximate $i_v^*$ for node $v \in \mathsf{V}$ via $\widehat{i_v^*} = \frac{1}{2}\left(\widetilde{i_v^*} + i_v^{*+}\right)$, where

$$\widetilde{i_v^*} = f_{(C, C_v)}(i^{*-}, i_v^{*+}, \deg(v)) = k_0 + k_1 i^{*-} + k_2 i_v^{*+} + k_3 \deg(v)$$

can be statistically derived from the data. Note that the heuristic function $\widehat{i_v^*}$ could be refined via more extensive numerical studies. We define the approximation error for network $\mathsf{G}$ as $\mathsf{err}_\mathsf{G} = \sum_{v \in \mathsf{V}}(\widehat{i_v^*} - i_v^*)$, because $\sum_{v \in \mathsf{V}} i_v^*$ is an important factor for cyber defense decision-making. For practical use, it is desired that $\mathsf{err}_\mathsf{G} \geq 0$, meaning that the defender never underestimates the threats, and at the same time $\mathsf{err}_\mathsf{G} \approx 0$, meaning that the defender does not overestimate the threats (i.e., does not overprovision defense resources) too much.

**ER networks.** For the ER network, we obtain the following formulas:

- For $(C, C_v)$=(Gaussian, Frank), we have $\widehat{i_v^*} = -0.01759 + 0.3142 i^{*-} + 0.7294 i_v^{*+} - 0.0002575 \deg(v)$.

- For $(C, C_v)$=(Gaussian, Clayton), we have $\widehat{i_v^*} = -0.0174076 + 0.3150585 i^{*-} + 0.7281992 i_v^{*+} - 0.0002596 \deg(v)$.

For $(C, C_v)$=(Gaussian, Frank), the average of the $\mathsf{err}_\mathsf{G}$'s over the 307 combinations of $(C, C_v, \alpha, \beta, \gamma)$ is 46, meaning that the approximation method only overestimates 46 infected nodes in an ER network of 1000 nodes. In comparison, the average of the $\sum_{v \in \mathsf{V}}(i_v^{*+} - i_v^*)$'s over the 307 combinations of $(C, C_v, \alpha, \beta, \gamma)$ is 93, meaning that the upper bound overestimates 93 infected nodes (i.e., the approximation method is indeed better); the average of the $\sum_{v \in \mathsf{V}}(i^- - i_v^*)$'s is -52.7, meaning that the lower bound underestimates 52.7 infected nodes in an ER network of 1000 nodes. Finally, we note that among the 307 combinations of $(C, C_v, \alpha, \beta, \gamma)$, the maximum $\mathsf{err}_\mathsf{G}$ is 165.2, which is elaborated in Figure 3(a) and will be discussed further, and the minimum $\mathsf{err}_\mathsf{G}$ is 4.1, which is elaborated in Figure 3(b) and will be discussed further as well. For $(C, C_v)$=(Gaussian, Clayton), the average of the $\mathsf{err}_\mathsf{G}$'s over the 307 combinations of $(C, C_v, \alpha, \beta, \gamma)$ is 46.5, meaning that the approximation method only overestimates 46.5 infected nodes in an ER network of 1000 nodes. In comparison, the average of the $\sum_{v \in \mathsf{V}}(i_v^{*+} - i_v^*)$'s over the 307 combinations of $(C, C_v, \alpha, \beta, \gamma)$ is 93, meaning that the upper bound overestimates 93 infected nodes in an ER network of 1000 nodes; the average of the 307 $\sum_{v \in \mathsf{V}}(i^- - i_v^*)$'s is -52.5, meaning that the lower bound underestimates 52.5 infected nodes in an ER network of 1000 nodes. Among the 307 instances, the maximum $\mathsf{err}_\mathsf{G}$ is 165.0, which is elaborated in Figure 3(d) and will be discussed further, and the minimum $\mathsf{err}_\mathsf{G}$ is 4.2, which is elaborated in Figure 3(e) and will be discussed further. In summary, cyber defense decision-making can be based on the approximation method, which takes advantage of the upper and lower bounds and would be better (smaller) than the upper bound.

As a side-product, we would like to highlight the phenomenon that the equilibrium infection probability $i_v^*$ increases with node degree $\deg(v)$. This phenomenon was observed in [19] in the absence of dependence, and persists in the presence of dependence as we elaborate below. We consider $i_v^{*+}, i_v^*, \widehat{i_v^*}$ and $i^{*-}$ with respect to distinct node degrees, by taking the average over the nodes of the same degree when needed. For $(C, C_v)$=(Gaussian,Frank), Figures 3(a)-3(b) plot the infection probabilities corresponding to the $(\alpha, \beta, \gamma)$ that leads to the maximum and minimum $\mathsf{err}_\mathsf{G}$, respectively; Figure 3(c) plots the infection probabilities averaged over the 307 combinations of $(\alpha, \beta, \gamma)$ that satisfy condition (8). For $(C, C_v)$=(Gaussian,Clayton), Figures 3(d)-3(e) plot the infection probabilities corresponding to the $(\alpha, \beta, \gamma)$ that leads to the maximum and minimum $\text{err}_\mathsf{G}$, respectively; Figure 4(f) plots the infection probabilities averaged over the 307 combinations of $(\alpha, \beta, \gamma)$ that satisfy condition (8). We observe that the approximation $\widehat{i_v^*}$ can slightly underestimate the infection probability $i_v^*$ for node $v$ of degree $\deg(v) \leq 5$, but the overall estimation $\sum_{v \in \mathsf{V}} \widehat{i_v^*}$ is still above the actual threats $\sum_{v \in \mathsf{V}} i_v^*$ (as mentioned above). More importantly, we observe that $i_v^*$ (solid curves) increases with $\deg(v)$. This hints that there might be some universal scaling laws, in the presence or absence of dependence. It is an interesting future work to identify the possible scaling law.

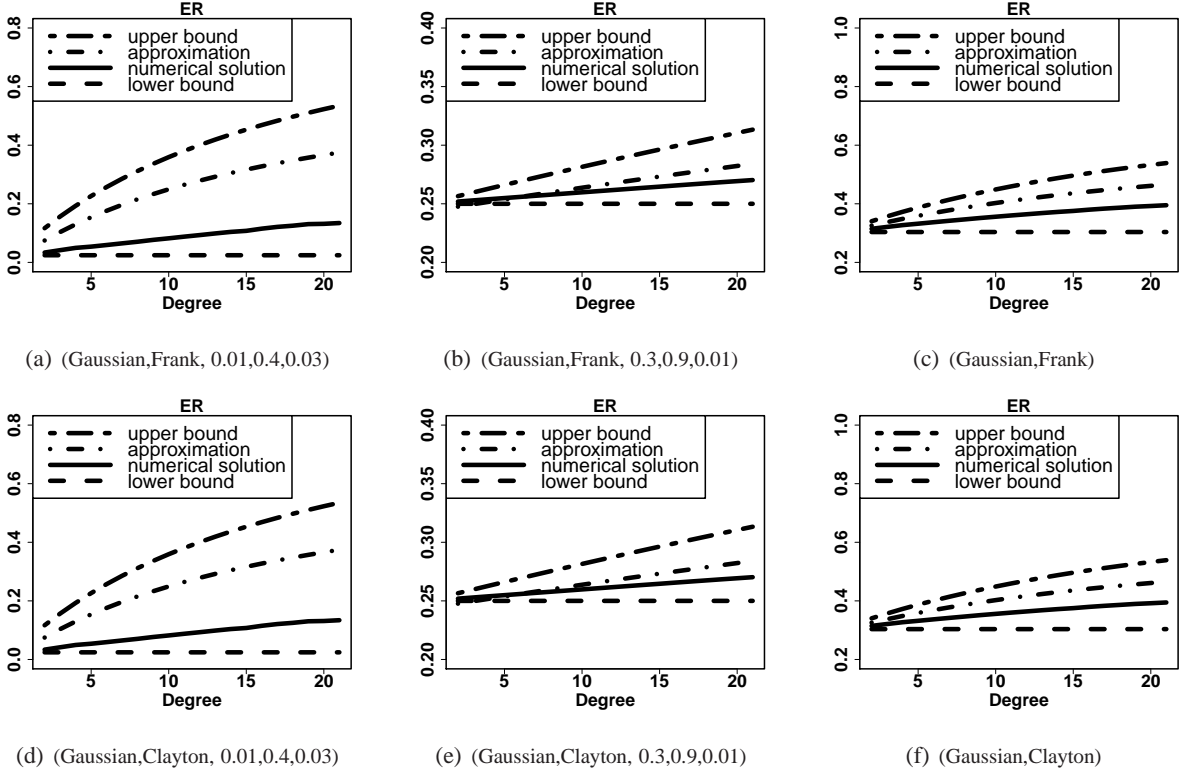

Figure 3: ER networks: upper bound vs. approximation vs. numerical solution vs. lower bound with respect to $(C, C_v, \alpha, \beta, \gamma)$ or $(C, C_v)$.

**Power-law networks.** For power-law networks, we obtain the following formulas in a similar fashion:

- For $(C, C_v)$=(Gaussian, Frank), we have $\widehat{i_v^*} = -0.007395 + 0.34705 i_v^{*-} + 0.67205 i_v^{*+} + 0.00013505 \deg(v)$.

- For $(C, C_v)$=(Gaussian, Clayton), we have $\widehat{i_v^*} = -0.007365 + 0.34765 i_v^{*-} + 0.6714 i_v^{*+} + 0.00013525 \deg(v)$.

For $(C, C_v)$=(Gaussian, Frank), the average of the $\text{err}_\mathsf{G}$'s over the 125 combinations of $(C, C_v, \alpha, \beta, \gamma)$ is 25, meaning that the approximation only overestimates 25 infected nodes in a power-law network of 1000 nodes. In comparison, the average of the $\sum_{v \in \mathsf{V}}(i_v^{*+} - i_v^*)$'s over the 125 combinations of $(C, C_v, \alpha, \beta, \gamma)$ is 50.8, meaning that the upper bound overestimates 50.8 infected nodes (i.e., the approximation method is better); the average of the $\sum_{v \in \mathsf{V}}(i^- - i_v^*)$'s is -26, meaning that the lower bound underestimates 26 infected nodes. Among the 125 combinations of $(C, C_v, \alpha, \beta, \gamma)$, the maximum $\text{err}_\mathsf{G}$ is 84.5, which is elaborated in Figure 4(a) and will be discussed further, and the minimum $\text{err}_\mathsf{G}$ is 7.1, which is elaborated in Figure 4(b). For $(C, C_v)$=(Gaussian, Clayton), the average of the $\text{err}_\mathsf{G}$'s over the 126 combinations of $(C, C_v, \alpha, \beta, \gamma)$ is 25.4, meaning that the approximation only overestimates 25.4 infected nodes in a power-law network of 1000 nodes. In comparison, the average of the $\sum_{v \in \mathsf{V}}(i_v^{*+} - i_v^*)$'s over the 126 combinations of $(C, C_v, \alpha, \beta, \gamma)$ is 50.8, meaning that the upper bound overestimates 50.8 infected nodes; the average of the 126 $\sum_{v \in \mathsf{V}}(i^- - i_v^*)$'s is -26, meaning that the lower bound underestimates 26 infected nodes. Among the 126 instances, the maximum $\text{err}_\mathsf{G}$ is 84.5, which is elaborated in Figure 4(d), and the minimum $\text{err}_\mathsf{G}$ is 7.2, which is elaborated in Figure 4(e). In summary, cyber defense decision-making can use the approximation method, which takes advantage of the upper and lower bounds.

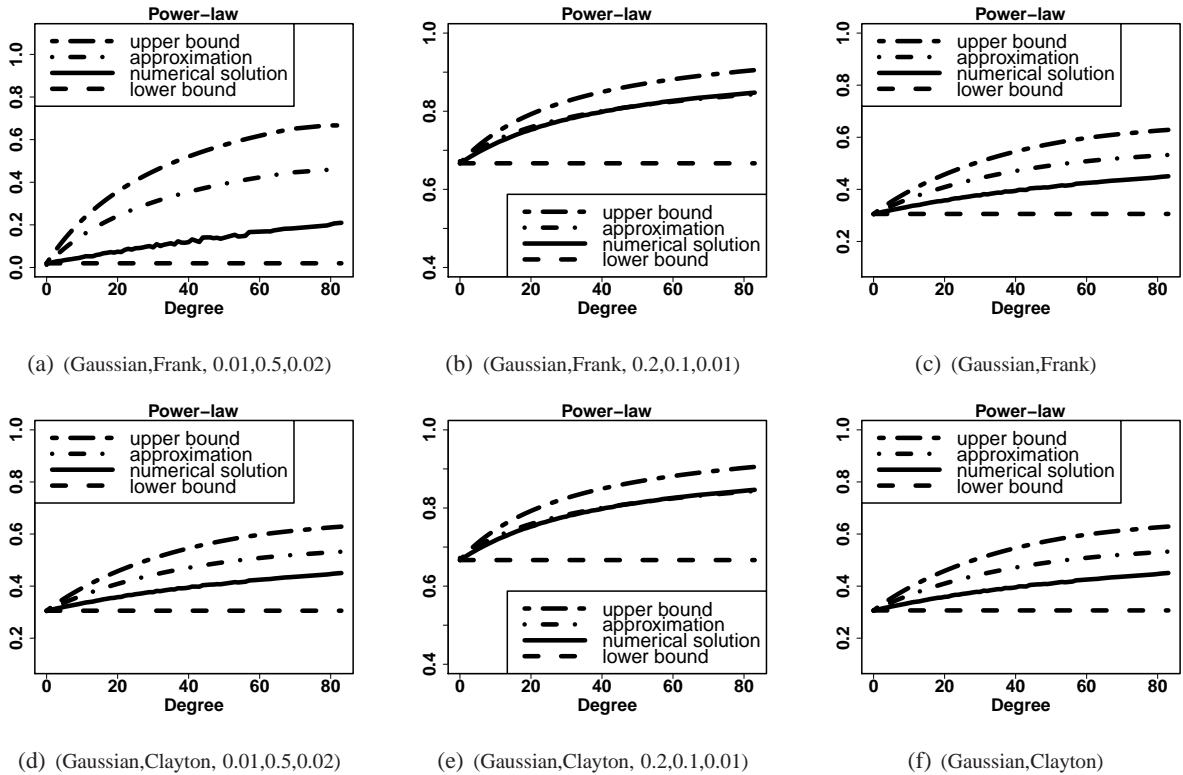

Figure 4: Power-law networks: upper bound vs. approximation vs. numerical solution vs. lower bound with respect to $(C, C_v, \alpha, \beta, \gamma)$ or $(C, C_v)$. In Figures 4(b) and 4(e), the approximation result matches the numerical solution almost perfectly.

We also would like to highlight the phenomenon that the equilibrium infection probability $i_v^*$ increases with node degree $\deg(v)$ in power-law networks. Similarly, for $(C, C_v)$=(Gaussian,Frank), Figures 4(a)-4(b) plot respectively the infection probabilities corresponding to the $(\alpha, \beta, \gamma)$ that leads to the maximum and minimum $\text{err}_G$, and Figure 4(c) plots the infection probabilities averaged over the 125 combinations of $(C, C_v, \alpha, \beta, \gamma)$ that satisfy condition (8). For $(C, C_v)$=(Gaussian,Clayton), Figures 4(a)-4(b) plot respectively the infection probabilities corresponding to the $(\alpha, \beta, \gamma)$ that leads to the maximum $\text{err}_G$, and Figure 4(c) plots the infection probabilities averaged over the 126 combinations of $(C, C_v, \alpha, \beta, \gamma)$ that satisfy condition (8). We observe that the approximation $\widehat{i_v^*}$ never underestimates the infection probability $i_v^*$ for any node $v$. We also observe that $i_v^*$ (solid curves) increases with $\deg(v)$, but exhibits a higher nonlinearity when compared with the ER networks.

### 3.5 Bounds for Non-Equilibrium Infection Probabilities

It is important to characterize the behavior of $i_v(t)$ even if it never enters any equilibrium. For this purpose, we want to seek some bounds for $i_v(t)$, no matter whether the system converges to an equilibrium or not. Such characterization is useful because, for example, the upper bound can be used for the worst-case scenario decision-making. It is worth mentioning that non-equilibrium states/behaviors are always hard to characterize.

**Proposition 5** (bounds for non-equilibrium probabilities) *Let $\overline{\lim}_{t \to \infty} i_v(t)$ and $\underline{\lim}_{t \to \infty} i_v(t)$ denote the upper and lower limits of $i_v(t)$, $v \in \mathsf{V}$. Then,*

$$i_v^- \leq \underline{\lim}_{t \to \infty} i_v(t) \leq \overline{\lim}_{t \to \infty} i_v(t) \leq i_v^+,$$

*where*

$$i_v^- = \begin{cases} \dfrac{1 - C(\delta_{C_v}(1-\gamma\nu), 1-\alpha)}{\beta + 1 - C(\delta_{C_v}(1-\gamma\nu), 1-\alpha)}, & C(\delta_{C_v}(1-\gamma\nu), 1-\alpha) \geq \beta, \\ [\beta - C(\delta_{C_v}(1-\gamma\nu), 1-\alpha)](1-\mu_v) + 1 - \beta, & otherwise, \end{cases}$$

*and*

$$i_v^+ = \begin{cases} \dfrac{1 - C\left(C_v\left(1-\gamma\mu_{v,1}, \ldots, 1-\gamma\mu_{v,\deg(v)}\right), 1-\alpha\right)}{\beta + 1 - C\left(C_v\left(1-\gamma\mu_{v,1}, \ldots, 1-\gamma\mu_{v,\deg(v)}\right), 1-\alpha\right)}, & C\left(C_v\left(1-\gamma\mu_{v,1}, \ldots, 1-\gamma\mu_{v,\deg(v)}\right), 1-\alpha\right) > \beta \\ \left[\beta - C\left(C_v\left(1-\gamma\mu_{v,1}, \ldots, 1-\gamma\mu_{v,\deg(v)}\right), 1-\alpha\right)\right](1-i_v^-) + 1 - \beta, & otherwise \end{cases}$$

*with* $\delta_{C_v}(1-\gamma\nu) = C_v(1-\gamma\nu, \ldots, 1-\gamma\nu)$, $\mu_v = \max\{1-\beta, \min\{\gamma\deg(v) + \alpha, 1\}\}$ *and* $\nu = \min\{1-\beta, \alpha\}$.

**Proof** By observing the monotonicity in Eq. (4), we note that $i_v(t) \geq \nu$ for all $v \in \mathsf{V}$. Replacing $i_{v,j}(t)$ with $\nu$ in Eq. (4) yields

$$\begin{aligned} i_v(t+1) &\geq (1-\beta)i_v(t) + [1 - C(\delta_{C_v}(1-\gamma\nu), 1-\alpha)](1 - i_v(t)) \\ &= [C(\delta_{C_v}(1-\gamma\nu), 1-\alpha) - \beta]i_v(t) + 1 - C(\delta_{C_v}(1-\gamma\nu), 1-\alpha). \end{aligned}$$

If $C(\delta_{C_v}(1-\gamma\nu), 1-\alpha) > \beta$, by taking the lower limit on both sides we obtain

$$\underline{\lim}_{t \to \infty} i_v(t+1) \geq \frac{1 - C(\delta_{C_v}(1-\gamma\nu), 1-\alpha)}{\beta + 1 - C(\delta_{C_v}(1-\gamma\nu), 1-\alpha)}.$$

If $C(\delta_{C_v}(1-\gamma\nu), 1-\alpha) \leq \beta$, we have

$$\begin{aligned} i_v(t+1) &\geq [C(\delta_{C_v}(1-\gamma\nu), 1-\alpha) - \beta]\mu_v + 1 - C(\delta_{C_v}(1-\gamma\nu), 1-\alpha) \\ &= [\beta - C(\delta_{C_v}(1-\gamma\nu), 1-\alpha)](1-\mu_v) + 1 - \beta. \end{aligned}$$

Hence, $\underline{\lim}_{t\to\infty} i_v(t) \geq i_v^-$.

For the upper bound, by applying Lemma 1 to Eq. (4) we have

$$\begin{aligned} i_v(t+1) &\leq (1-\beta)i_v(t) + [1 - \max\{\max\{2 - \gamma\deg(v) - \alpha, 1-\alpha\} - 1, 0\}](1 - i_v(t)) \\ &= (1-\beta)i_v(t) + \min\{\gamma\deg(v) + \alpha, 1\}(1 - i_v(t)) \\ &\leq \max\{1-\beta, \min\{\gamma\deg(v) + \alpha, 1\}\} = \mu_v. \end{aligned} \quad (15)$$

By replacing $i_{v,j}$ with $\mu_{v,j}$'s in Eq. (4) yields

$$\begin{aligned} i_v(t+1) &\leq (1-\beta)i_v(t) + \left[1 - C\left(C_v\left(1-\gamma\mu_{v,1}, \ldots, 1-\gamma\mu_{v,\deg(v)}\right), 1-\alpha\right)\right](1 - i_v(t)) \\ &= \left[C\left(C_v\left(1-\gamma\mu_{v,1}, \ldots, 1-\gamma\mu_{v,\deg(v)}\right), 1-\alpha\right) - \beta\right]i_v(t) \\ &\quad + 1 - C\left(C_v\left(1-\gamma\mu_{v,1}, \ldots, 1-\gamma\mu_{v,\deg(v)}\right), 1-\alpha\right). \end{aligned}$$

If $C\left(C_v\left(1-\gamma\mu_{v,1}, \ldots, 1-\gamma\mu_{v,\deg(v)}\right), 1-\alpha\right) > \beta$, then

$$\overline{\lim}_{t\to\infty} \leq \frac{1 - C\left(C_v\left(1-\gamma\mu_{v,1}, \ldots, 1-\gamma\mu_{v,\deg(v)}\right), 1-\alpha\right)}{\beta + 1 - C\left(C_v\left(1-\gamma\mu_{v,1}, \ldots, 1-\gamma\mu_{v,\deg(v)}\right), 1-\alpha\right)}.$$

If $C\left(C_v\left(1-\gamma\mu_{v,1}, \ldots, 1-\gamma\mu_{v,\deg(v)}\right), 1-\alpha\right) \leq \beta$, then we have

$$\begin{aligned} \overline{\lim}_{t\to\infty} i_v(t+1) &\leq \overline{\lim}_{t\to\infty}\left\{\left[C\left(C_v\left(1-\gamma\mu_{v,1}, \ldots, 1-\gamma\mu_{v,\deg(v)}\right), 1-\alpha\right) - \beta\right]i_v(t) \right. \\ &\qquad\qquad \left. + 1 - C\left(C_v\left(1-\gamma\mu_{v,1}, \ldots, 1-\gamma\mu_{v,\deg(v)}\right), 1-\alpha\right)\right\} \\ &\leq \left[C\left(C_v\left(1-\gamma\mu_{v,1}, \ldots, 1-\gamma\mu_{v,\deg(v)}\right), 1-\alpha\right) - \beta\right]\underline{\lim}_{t\to\infty} i_v(t) \\ &\qquad + 1 - C\left(C_v\left(1-\gamma\mu_{v,1}, \ldots, 1-\gamma\mu_{v,\deg(v)}\right), 1-\alpha\right) \\ &\leq \left[C\left(C_v\left(1-\gamma\mu_{v,1}, \ldots, 1-\gamma\mu_{v,\deg(v)}\right), 1-\alpha\right) - \beta\right] i_v^- \\ &\qquad + 1 - C\left(C_v\left(1-\gamma\mu_{v,1}, \ldots, 1-\gamma\mu_{v,\deg(v)}\right), 1-\alpha\right). \end{aligned}$$

Hence, we have $\overline{\lim}_{t\to\infty} i_v(t+1) \le i_v^+$. ∎

**When are the bounds tight?** It is important to know when the bounds are tight because the defender can use the upper bound $i_v^+$ for decision-making, especially when the spreading never enters any equilibrium. Note that when $\gamma \ll 1$, it holds that

$$C(\delta_{C_v}(1-\gamma\nu), 1-\alpha) \approx C(1, 1-\alpha) = 1-\alpha,$$

and

$$C\left(C_v\left(1-\gamma\mu_{v,1}, \ldots, 1-\gamma\mu_{v,\deg(v)}\right), 1-\alpha\right) \approx C\left(C_v(1,\ldots,1), 1-\alpha\right) = 1-\alpha.$$

Therefore, in the case $\gamma \ll 1$ and $\alpha + \beta < 1$, we have

$$i_v^- \approx \frac{1 - C(\delta_{C_v}(1-\gamma\nu), 1-\alpha)}{\beta + 1 - C(\delta_{C_v}(1-\gamma\nu), 1-\alpha)} \approx \frac{\alpha}{\beta+\alpha},$$

$$i_v^+ \approx \frac{1 - C\left(C_v\left(1-\gamma\mu_{v,1},\ldots,1-\gamma\mu_{v,\deg(v)}\right), 1-\alpha\right)}{\beta + 1 - C\left(C_v\left(1-\gamma\mu_{v,1},\ldots,1-\gamma\mu_{v,\deg(v)}\right), 1-\alpha\right)} \approx \frac{\alpha}{\beta+\alpha}.$$

This means that the bounds are tight when the attack-power is not strong.

In the case $\gamma \deg(v) \ll 1$ and $\alpha + \beta \ge 1$, we can similarly have

$$i_v^- \approx \alpha(2-\alpha-\beta), \quad i_v^+ \approx (\beta+\alpha-1)\left[1-\alpha(2-\alpha-\beta)\right] + 1 - \beta.$$

Therefore, the difference between the upper bound and lower bound is

$$i_v^+ - i_v^- \approx \alpha(\alpha+\beta-1)^2.$$

Therefore, the bounds are tight when $(\alpha+\beta)$ is not far from 1 or $\alpha$ is close to zero.

**Are the equilibrium bounds always tighter than the non-equilibrium bounds?** We observe the following: under the condition $\gamma \deg(v) \ll 1$, we have $i^{*-} \approx i_v^{*+} \approx \alpha/(\alpha+\beta)$; under the condition $\gamma \deg(v) \ll 1$ *and* the condition $\alpha + \beta < 1$, we have $i_v^- \approx i_v^+ \approx \alpha/(\alpha+\beta)$. This means that the equilibrium bounds are widely applicable than the same non-equilibrium bounds, namely that the equilibrium bounds are strictly tighter than the non-equilibrium bounds.

## 4 Side-Effects of Assuming Away the Dependences

In the above we have characterized epidemic equilibrium thresholds, equilibrium infection probabilities, and non-equilibrium infection probabilities while accommodating arbitrary dependences. In order to characterize the side-effects of assuming away the dependences, we consider the degree of dependences as captured by the *concordance order* between copulas (reviewed in Section 2). In order to draw cyber security insights at a higher level of abstraction, we also consider three kinds of qualitative dependences: *positive dependence*, *independence* and *negative dependence*, whose degrees of dependence are in decreasing order. Specifically, positive (negative) dependence between the push-based attacks means

$$1 - C_v\left(1-\gamma i_{v,1}, \ldots, 1-\gamma i_{v,\deg(v)}\right) \ge (\le) 1 - \prod_{j=1}^{\deg(v)} \left(1-\gamma i_{v,j}\right),$$

and positive (negative) dependence between the push-based attacks and the pull-based attacks means

$$1 - C\left(C_v\left(1-\gamma i_{v,1}(t), \ldots, 1-\gamma i_{v,\deg(v)}(t)\right), 1-\alpha\right) \ge (\le) 1 - (1-\alpha)C_v\left(1-\gamma i_{v,1}(t), \ldots, 1-\gamma i_{v,\deg(v)}(t)\right),$$

where equality means independence. To simplify the notations, let pd stand for *positive dependence*, ind stand for *independence*, and nd stand for *negative dependence*. Let $x \in \{\text{pd}, \text{ind}, \text{nd}\}$ denote the dependence structure between the push-based attacks and the pull-based attacks, as captured by copula $C$. Let $y \in \{\text{pd}, \text{ind}, \text{nd}\}$ denote the dependence structure between the push-based attacks, as captured by copula $C_v$. Therefore, the dependence structures can be represented by a pair $(x, y)$.

## 4.1 Side-Effects on Equilibrium Infection Probabilities and Thresholds

For fixed $\mathsf{G} = (\mathsf{V}, \mathsf{E}), \alpha, \beta, \gamma$, we compare the effects of two groups of dependences (i.e., copulas) $\{C, C_v, v \in \mathsf{V}\}$ and $\{C', C'_v, v \in \mathsf{V}\}$. Corresponding to the two groups of copulas, we denote by $i_v(t)$ and $i'_v(t)$ the respective infection probabilities of node $v \in \mathsf{V}$ at time $t \geq 0$. Let $i^*_{v,\mathsf{x},\mathsf{y}}$ denote the equilibrium infection probability of node $v$, namely $i^*_v$, under dependence structure $(\mathsf{x}, \mathsf{y})$.

**Side-effects on the equilibrium infection probabilities.** We present a result about the impact of the dependence structures on the equilibrium infection probabilities. This result will allow us to derive the side-effects of assuming away the dependences.

**Proposition 6** (comparison between the effects of different dependence structures on equilibrium infection probabilities) *Suppose the condition underlying Lemma 3 holds, namely $\rho(A) < \dfrac{(\beta + \alpha)^2}{\gamma \beta}$ so that system (4) has a unique equilibrium. If for all $v \in \mathsf{V}$, we have*

$$C\left(C_v\left(u_1, \ldots, u_{\deg(v)}\right), u_0\right) \leq C'\left(C'_v\left(u_1, \ldots, u_{\deg(v)}\right), u_0\right), \tag{16}$$

*where $0 \leq u_j \leq 1$ for $j = 0, \ldots, \deg(v)$, then we have $\mathbf{i}^* \geq \mathbf{i}'^*$.*

**Proof** Note that $\mathbf{i}^*$ and $\mathbf{i}'^*$ are respectively the unique positive solutions of $\boldsymbol{f}(\mathbf{i}^*) = \mathbf{0}$ and $\boldsymbol{g}(\mathbf{i}'^*) = \mathbf{0}$, where $\boldsymbol{f} = (f_1, \ldots, f_N)$ and $\boldsymbol{g} = (g_1, \ldots, g_N)$ with

$$\begin{aligned} f_v(\mathbf{i}) &= \left[1 - C\left(C_v\left(1 - \gamma i_{v,1}, \ldots, 1 - \gamma i_{v,\deg(v)}\right), 1 - \alpha\right)\right](1 - i_v) - \beta i_v, \quad v \in \mathsf{V} \\ g_v(\mathbf{i}) &= \left[1 - C'\left(C'_v\left(1 - \gamma i_{v,1}, \ldots, 1 - \gamma i_{v,\deg(v)}\right), 1 - \alpha\right)\right](1 - i_v) - \beta i_v, \quad v \in \mathsf{V}. \end{aligned}$$

Since $\boldsymbol{f}(\mathbf{0}) = \boldsymbol{g}(\mathbf{0}) = \boldsymbol{\alpha} > \mathbf{0}$ and $\boldsymbol{f} \geq \boldsymbol{g}$, we have $\boldsymbol{g}(\mathbf{i}^*) \leq \boldsymbol{f}(\mathbf{i}^*) = \mathbf{0}$. Since both $f$ and $g$ are continuous, we have $\mathbf{i}'^* \leq \mathbf{i}^*$. ∎

The cyber security insights/implications of Proposition 6 is: The stronger the negative (positive) dependences between the attack events, the lower (higher) the equilibrium infection probabilities. More specifically, we have $i^*_{v,\mathsf{pd},\mathsf{y}} \geq i^*_{v,\mathsf{ind},\mathsf{y}} \geq i^*_{v,\mathsf{nd},\mathsf{y}}$ for any $\mathsf{y} \in \{\mathsf{pd}, \mathsf{ind}, \mathsf{nd}\}$ and $i^*_{v,\mathsf{x},\mathsf{pd}} \geq i^*_{v,\mathsf{x},\mathsf{ind}} \geq i^*_{v,\mathsf{x},\mathsf{nd}}$ for any $\mathsf{x} \in \{\mathsf{pd}, \mathsf{ind}, \mathsf{nd}\}$. Therefore, the side-effects of assuming away the dependences between attack events are: If the positive (negative) dependence is assumed away, the resulting equilibrium infection probability underestimates (overestimates) the actual equilibrium infection probability. This means the following: when the positive dependence between attack events is assumed away, the cyber defense decisions based on $i^*_{v,\mathsf{ind},\mathsf{ind}}$ ($< i^*_{v,\mathsf{pd},\mathsf{pd}}$) can render the deployed defense useless; when the negative dependence is assumed away between attack events, the cyber defense decisions based on $i^*_{v,\mathsf{ind},\mathsf{ind}}$ ($> i^*_{v,\mathsf{nd},\mathsf{nd}}$) can waste defense resources. We will use numerical examples below to confirm these insights. Another important insight is: if the defender can seek to impose negative dependence on the cyber attacks, the cyber defense effect is better of. We believe that this insight will shed light on research of future cyber defense mechanisms, and highlights the value of theoretical studies in terms of their practical guidance.

**Side-effects on the epidemic equilibrium threshold.** Corollary 1 gives a sufficient condition under which the epidemic spreading enters the equilibrium. Here we define

$$\tau \stackrel{def}{=} \min\left\{\frac{1 - \max_{v \in V}|1 - \beta/(1 - i^*_v)|}{\gamma}, \frac{(\beta + \alpha)^2}{\gamma \beta}\right\}, \tag{17}$$

with respect to a group of copulas $\{C, C_v, v \in \mathsf{V}\}$. According to Eq. (8), $\rho(A) \leq \tau$ means that the epidemic spreading converges to the equilibrium. Similarly, we can define $\tau'$ with respect to another group of copulas $\{C', C'_v, v \in \mathsf{V}\}$. We want to compare $\tau$ and $\tau'$ with respect to the relation between $\{C, C_v, v \in \mathsf{V}\}$ and $\{C', C'_v, v \in \mathsf{V}\}$.

**Proposition 7** *Under the conditions of Proposition 6, namely $\rho(A) < \frac{(\beta+\alpha)^2}{\gamma\beta}$ so that system (4) has a unique equilibrium and $C\left(C_v\left(u_1,\ldots,u_{\deg(v)}\right), u_0\right) \leq C'\left(C'_v\left(u_1,\ldots,u_{\deg(v)}\right), u_0\right)$ for all $v \in \mathsf{V}$, we have*

(i) *if $1-\beta \leq i^{*-}$, then $\tau \leq \tau'$;*

(ii) *if $1-\beta \geq i^{*+}$, then $\tau \geq \tau'$,*

*where $i^{*+} \stackrel{def}{=} \max_{v\in\mathsf{V}} i_v^{*+}$, $i^{*-}$ and $i_v^{*+}$ are defined in Proposition 1.*

**Proof** According to Proposition 1, we know that $i^{*-} \leq i_v^* \leq i^{*+}$, which implies $\frac{\beta}{1-i^{*-}} \leq \frac{\beta}{1-i_v^*} \leq \frac{\beta}{1-i^{*+}}$. According to Eq. (17), $\tau$ is decreasing in $\max_{v\in V} |1-\frac{\beta}{1-i_v^*}|$. Therefore, $\tau$ is decreasing in $i_v^*$ when $1-\beta \leq i^{*-}$, and increasing in $i_v^*$ when $1-\beta \geq i^{*+}$. By Proposition 6, we get the desired results. ∎

In order to draw insights while simplifying the discussion, let $\tau_{\mathsf{x},\mathsf{y}}$ denote the $\tau$ as defined in Eq. (17) with respect to dependence structures $(\mathsf{x},\mathsf{y})$. The cyber security implication of Proposition 7 is: First, under some circumstances, the stronger the dependences between the cyber attacks, the more restrictive the epidemic equilibrium threshold. More specifically, under the condition $1-\beta \leq i^{*-}$, we have for all $v \in \mathsf{V}$:

$$\tau_{\mathsf{nd},\mathsf{y}} \geq \tau_{\mathsf{ind},\mathsf{y}} \geq \tau_{\mathsf{pd},\mathsf{y}} \text{ and } \tau_{\mathsf{x},\mathsf{nd}} \geq \tau_{\mathsf{x},\mathsf{ind}} \geq \tau_{\mathsf{x},\mathsf{pd}}.$$

This means that under the above circumstances, assuming away the positive dependences between the attacks will lead to incorrect epidemic equilibrium threshold, and assuming away the negative dependences between the make the epidemic equilibrium threshold unnecessarily restrictive. This further highlights the value for the defender to render the dependences negative, provided that $1-\beta \leq i^{*-}$.

Second, under certain other circumstances, the stronger the dependences, the less restrictive the epidemic equilibrium threshold. More specifically, under the condition $1-\beta \geq i^{*+}$, we have

$$\tau_{\mathsf{nd},\mathsf{y}} \leq \tau_{\mathsf{ind},\mathsf{y}} \leq \tau_{\mathsf{pd},\mathsf{y}} \text{ and } \tau_{\mathsf{x},\mathsf{nd}} \leq \tau_{\mathsf{x},\mathsf{ind}} \leq \tau_{\mathsf{x},\mathsf{pd}}.$$

This means that assuming away the negative dependences between the attacks will lead to incorrect epidemic equilibrium threshold, and assuming away the positive dependences will make the epidemic equilibrium threshold unnecessarily restrictive. Moreover, while rendering the dependences negative can lead to smaller equilibrium infection probabilities, it imposes a very restrictive epidemic equilibrium threshold when $1-\beta \geq i^{*+}$. This means that when applying the above insights to guide practice, the defender must be aware of the parameter regions corresponding to the cyber security posture.

**Numerical examples.** In order to illustrate the above analytic results, we consider the example of star network with $N=11$ nodes. We assume that the dependence between the push-based and the pull-based attacks can be captured by the Gaussian copula $C$ with parameter $\sigma$ and the dependence between the push-based attacks launched from the leaves against the hub can be captured by copula $C_v$, which is the Clayton copula with parameter $\theta$. These two copulas are reviewed in Section 2. We consider two sets of parameters $(\alpha,\beta,\gamma) = (0.2, 0.5, 0.05)$ and $(\alpha,\beta,\gamma) = (0.4, 0.7, 0.05)$. From Eqs. (11) and (12), we can compute the equilibrium infection probabilities $i_h^*$ for the hub and $i_l^*$ for the leaves, and the threshold $\tau$ as defined in (17). Note that the copulas are increasing in their parameters in the concordance order. By Proposition 6, both $i_h^*$ and $i_l^*$ are decreasing in $\theta$ ($\sigma$) given $\sigma$ ($\theta$), as confirmed by Tables 1-2. Note that for star networks, the condition $1-\beta \geq i^{*+}$ in Proposition 7 can be relaxed as $1-\beta \geq i_h^{*+}$, where $i_h^{*+}$ is defined in Proposition 3. When $(\alpha,\beta,\gamma) = (0.2, 0.5, 0.05)$, it is easy to verify $1-\beta \geq i_h^{*+}$, meaning that $\tau$ is decreasing in $\theta$ ($\sigma$) for fixed $\sigma$ ($\theta$). This is confirmed in Table 1. When $(\alpha,\beta,\gamma) = (0.4, 0.7, 0.05)$, the condition $1-\beta \leq i^{*-}$ in Proposition 7 is satisfied, meaning that $\tau$ is increasing in $\theta$ ($\sigma$) for fixed $\sigma$ ($\theta$). This is confirmed in Table 2. These examples also confirm the conclusion $i_h^* \geq i_l^*$ given by Proposition 2.

| $\theta$ | $\sigma = 0.5$ (nd) | | | $\sigma = 0$ (ind) | | | $\sigma = -0.5$ (pd) | | |
|---|---|---|---|---|---|---|---|---|---|
| | $i_h^*$ | $i_l^*$ | $\tau$ | $i_h^*$ | $i_l^*$ | $\tau$ | $i_h^*$ | $i_l^*$ | $\tau$ |
| 1.0 | .35 | .29 | 14.11 | .38 | .30 | 14.31 | .40 | .31 | 14.40 |
| 1.5 | .35 | .29 | 14.11 | .38 | .30 | 14.30 | .40 | .31 | 14.39 |
| 2.0 | .35 | .29 | 14.11 | .38 | .30 | 14.30 | .39 | .31 | 14.39 |
| 2.5 | .34 | .29 | 14.11 | .38 | .30 | 14.30 | .39 | .31 | 14.39 |
| 3.0 | .34 | .29 | 14.11 | .37 | .30 | 14.30 | .39 | .30 | 14.39 |
| 3.5 | .34 | .29 | 14.11 | .37 | .30 | 14.30 | .39 | .30 | 14.38 |
| 4.0 | .34 | .29 | 14.11 | .37 | .30 | 14.30 | .39 | .30 | 14.38 |
| 4.5 | .34 | .29 | 14.11 | .37 | .30 | 14.29 | .38 | .30 | 14.38 |
| 5.0 | .34 | .29 | 14.11 | .37 | .30 | 14.29 | .38 | .30 | 14.38 |
| 5.5 | .34 | .29 | 14.11 | .37 | .30 | 14.29 | .38 | .30 | 14.38 |
| 6.0 | .33 | .29 | 14.11 | .36 | .30 | 14.29 | .38 | .30 | 14.38 |

Table 1: $(\alpha, \beta, \gamma) = (0.2, 0.5, 0.05)$

| $\theta$ | $\sigma = 0.5$ (nd) | | | $\sigma = 0$ (ind) | | | $\sigma = -0.5$ (pd) | | |
|---|---|---|---|---|---|---|---|---|---|
| | $i_h^*$ | $i_l^*$ | $\tau$ | $i_h^*$ | $i_l^*$ | $\tau$ | $i_h^*$ | $i_l^*$ | $\tau$ |
| 1.0 | .39 | .37 | 17.11 | .41 | .37 | 16.09 | .44 | .38 | 15.20 |
| 1.5 | .39 | .37 | 17.15 | .41 | .37 | 16.16 | .43 | .38 | 15.29 |
| 2.0 | .39 | .37 | 17.18 | .41 | .37 | 16.21 | .43 | .38 | 15.36 |
| 2.5 | .39 | .37 | 17.21 | .41 | .37 | 16.26 | .43 | .38 | 15.43 |
| 3.0 | .38 | .37 | 17.24 | .41 | .37 | 16.31 | .43 | .38 | 15.50 |
| 3.5 | .38 | .37 | 17.27 | .41 | .37 | 16.35 | .43 | .38 | 15.56 |
| 4.0 | .38 | .37 | 17.30 | .41 | .37 | 16.39 | .43 | .38 | 15.62 |
| 4.5 | .38 | .37 | 17.31 | .41 | .37 | 16.43 | .42 | .38 | 15.67 |
| 5.0 | .38 | .37 | 17.33 | .41 | .37 | 16.47 | .42 | .38 | 15.72 |
| 5.5 | .38 | .37 | 17.35 | .40 | .37 | 16.50 | .42 | .38 | 15.77 |
| 6.0 | .38 | .37 | 17.37 | .40 | .37 | 16.53 | .42 | .38 | 15.81 |

Table 2: $(\alpha, \beta, \gamma) = (0.4, 0.7, 0.05)$

## 4.2 Side-Effects on the Non-Equilibrium Infection Probabilities

We now investigate the side-effects on the non-equilibrium infection probabilities $\mathbf{i}(t) = (i_1(t), \ldots, i_N(t))$, no matter whether the epidemic spreading converges to equilibrium or not.

**Proposition 8** (side-effects on the non-equilibrium infection probabilities) *Consider two vectors of infection probabilities $\mathbf{i}(t_0) \geq \mathbf{i}'(t_0)$ at some time $t_0 \geq 0$. Let $\mu = \max_{v \in \mathsf{V}} \mu_v = \max\{1 - \beta, \min\{\alpha + \gamma \mathrm{Deg}, 1\}\}$, where $\mathrm{Deg} = \max_{v \in \mathsf{V}} \deg(v)$. If condition (16) holds and*

$$\min_{v \in \mathsf{V}} \{C\left(\delta_{C_v}\left(1 - \gamma \mu\right), 1 - \alpha\right)\} \geq \beta, \tag{18}$$

*then $\mathbf{i}(t) \geq \mathbf{i}'(t)$ for all $t \geq t_0$.*

**Proof** We need to show that $\mathbf{i}(t+1) \geq \mathbf{i}'(t+1)$ when $\mathbf{i}(t) \geq \mathbf{i}'(t)$ is given. Note that

$$\begin{aligned}
i_v(t+1) &= \left[C\left(C_v\left(1 - \gamma i_{v,1}(t), \ldots, 1 - \gamma i_{v,\deg(v)}(t)\right), 1 - \alpha\right) - \beta\right](i_v(t) - 1) + 1 - \beta, \\
i'_v(t+1) &= \left[C'\left(C'_v\left(1 - \gamma i'_{v,1}(t), \ldots, 1 - \gamma i'_{v,\deg(v)}(t)\right), 1 - \alpha\right) - \beta\right](i'_v(t) - 1) + 1 - \beta.
\end{aligned}$$

According to Ineq. (15) in the proof of Proposition 5, we have $i_v(t) \leq \mu$ for all $v \in \mathsf{V}$. Then, conditions (16) and (18) imply

$$\begin{aligned}
C\left(C_v\left(1 - \gamma i_{v,1}(t), \ldots, 1 - \gamma i_{v,\deg(v)}(t)\right), 1 - \alpha\right) - \beta &\geq 0, \\
C'\left(C'_v\left(1 - \gamma i_{v,1}(t), \ldots, 1 - \gamma i_{v,\deg(v)}(t)\right), 1 - \alpha\right) - \beta &\geq 0.
\end{aligned}$$

Since $\mathbf{i}(t) \geq \mathbf{i}'(t)$, we have

$$\begin{aligned}
i_v(t+1) &\geq \left[C\left(C_v\left(1 - \gamma i_{v,1}(t), \ldots, 1 - \gamma i_{v,\deg(v)}(t)\right), 1 - \alpha\right) - \beta\right](i'_v(t) - 1) + 1 - \beta \\
&\geq \left[C'\left(C'_v\left(1 - \gamma i_{v,1}(t), \ldots, 1 - \gamma i_{v,\deg(v)}(t)\right), 1 - \alpha\right) - \beta\right](i'_v(t) - 1) + 1 - \beta \\
&\geq \left[C'\left(C'_v\left(1 - \gamma i'_{v,1}(t), \ldots, 1 - \gamma i'_{v,\deg(v)}(t)\right), 1 - \alpha\right) - \beta\right](i'_v(t) - 1) + 1 - \beta \\
&= i'_v(t+1).
\end{aligned}$$

Since the above holds for all $v \in \mathsf{V}$, we obtain the desired result. ∎

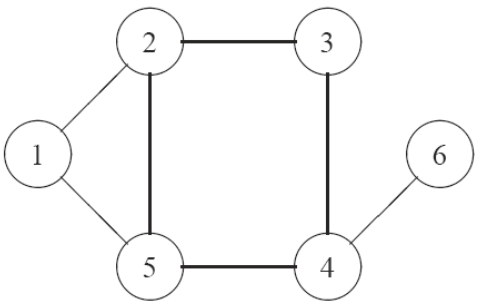

| | $t=6$ | | $t=7$ | | $t=8$ | |
|---|---|---|---|---|---|---|
| | $i_v(t)$ | $i'_v(t)$ | $i_v(t)$ | $i'_v(t)$ | $i_v(t)$ | $i'_v(t)$ |
| $v=1$ | 0.61 | 0.60 | 0.42 | 0.42 | 0.57 | 0.56 |
| 2 | 0.64 | 0.63 | **0.39** | **0.40** | 0.60 | 0.58 |
| 3 | **0.56** | **0.57** | 0.46 | 0.45 | 0.54 | 0.54 |
| 4 | 0.57 | 0.57 | 0.45 | 0.45 | 0.55 | 0.54 |
| 5 | **0.46** | **0.47** | 0.54 | 0.53 | **0.47** | **0.48** |
| 6 | 0.60 | 0.60 | 0.42 | 0.42 | 0.56 | 0.56 |

Figure 5: Clayton copulas with $(\theta, \eta) = (1, 1.5)$, $(\theta, \eta) = (10, 15)$, $(\alpha, \beta, \gamma) = (0.9, 0.9, 0.8)$.

One may wonder if a more succinct result than Proposition 8 could be obtained by, for example, eliminating condition (18). Here we use an example to show that if we eliminate condition (18), then Proposition 8 may not hold. Specifically, consider the network with six nodes illustrated in Figure 5. Suppose $C$ and the $C_v$'s for $v \in \mathsf{V}$ are Clayton copulas with positive parameters $\theta$ and $\eta$, namely

$$C(u_1, u_2) = \left[u_1^{-\theta} + u_2^{-\theta} - 1\right]^{-1/\theta} \quad \text{and} \quad C_v\left(u_1, \ldots, u_{\deg(v)}\right) = \left[\sum_{i=1}^{\deg(v)} u_i^{-\eta} - \deg(v) + 1\right]^{-1/\eta}.$$

Consider two groups of Clayton copulas respectively with parameters $(\theta, \eta) = (1, 1.5)$ and $(\theta, \eta) = (10, 15)$. Denote the corresponding infection probabilities by $i_v(t)$ and $i'_v(t)$, respectively. Set $(\alpha, \beta, \gamma) = (0.9, 0.9, 0.8)$, and in this case condition (18) is not satisfied. Set the initial infection probabilities as $\mathbf{i}(0) = \mathbf{i}'(0) = (0.2, 0.1, 0.3, 0.3, 0.6, 0.2)$. The table in Figure 5 shows $\mathbf{i}(t)$ and $\mathbf{i}'(t)$ for $t = 6, 7, 8$, from which we observe that $\mathbf{i}(t) \geq \mathbf{i}'(t)$ does not hold. This means that we cannot eliminate condition (18) in Proposition 8.

## 5 Conclusions

We have presented the first systematic investigation of cyber epidemic models with dependences. We have derived epidemic equilibrium thresholds, bounds for equilibrium infection probabilities, and bounds for non-equilibrium infection probabilities, while accommodating arbitrary dependences between the push-based attacks and the pull-based attacks as well as the dependences between the push-based attacks. In particular, we showed that assuming away the due dependences can render the results thereof unnecessarily restrictive or even incorrect.

Our study brings up a range of interesting research problems for further work. First, our characterization study assumes that the dependence or copula structures are given. It is important to know which dependence structures are more relevant than the others in practice. Second, it is ideal to obtain closed-form results on the equilibrium infection probabilities and the non-equilibrium infection probabilities. Third, if we cannot derive closed-form results for the (non-)equilibrium infection probabilities, it is important to seek bounds for these probabilities and systematically analyze their tightness.

**Acknowledgement**. This work was supported in part by ARO Grants # W911NF-12-1-0286 and # W911NF-13-1-0141, and by AFOSR Grant # FA9550-09-1-0165.